\documentclass[usegraphicx]{mn2e}
\newcommand\etal{et~al.~}

\newcommand\kms{{\,km\,s$^{-1}$}}

\usepackage{longtable}
\usepackage{deluxetable}
\begin{document}

\title
[NGC\,5044 group stellar populations]
{The anatomy of the NGC\,5044 group -- II. Stellar populations and star-formation histories}

\author[Mendel \etal]
{J. Trevor Mendel$^1$\thanks{tmendel@swin.edu.au}, Robert N. Proctor$^1$, Jesper Rasmussen$^2$, Sarah Brough$^1$,\and 
Duncan A. Forbes$^1$\\
\\
$^1$Centre for Astrophysics \& Supercomputing, Swinburne University, Mail H39, Hawthorn, VIC 3122,
Australia\\
$^2$Observatories of the Carnegie Institution (Chandra Fellow), 813 Santa Barbara Street, Pasadena, CA 91101, USA }

\maketitle

\begin{abstract}

The distribution of galaxy properties in groups and clusters holds important information on galaxy
evolution and growth of structure in the Universe.  While clusters have received appreciable
attention in this regard, the role of groups as fundamental to formation of the present day galaxy
population has remained relatively unaddressed.  Here we present stellar ages, metallicities and
$\alpha$-element abundances derived using Lick indices for 67 spectroscopically confirmed members of
the NGC\,5044 galaxy group with the aim of shedding light on galaxy evolution in the context of the
group environment. 
 
We find that galaxies in the NGC\,5044 group show evidence for a strong relationship between stellar
mass and metallicity, consistent with their counterparts in both higher and lower mass groups and
clusters.  Galaxies show no clear trend of age or $\alpha$-element abundance with mass, but
these data form a tight sequence when fit simultaneously in age, metallicity and stellar
mass.  In the context of the group environment, our data support the tidal disruption of low-mass
galaxies at small group-centric radii, as evident from an apparent lack of galaxies below
$\sim10^9\,M_{\odot}$ within $\sim100$\,kpc of the brightest group galaxy.  Using a joint analysis
of absorption- {\it and} emission-line metallicities, we are able to show that the star-forming
galaxy population in the NGC\,5044 group appears to require gas removal to explain the
$\sim$1.5\,dex offset between absorption- and emission-line metallicities observed in some cases.  A
comparison with other stellar population properties suggests that this gas removal is dominated by
galaxy interactions with the hot intragroup medium.

\end{abstract}


\section{Introduction}

Large photometric surveys have established that the observable galaxy population forms two distinct
sequences in colour-magnitude space: a red sequence of predominantly early-type galaxies and a
star-forming, late-type dominated blue sequence (e.g. Hogg \etal 2002; Blanton \etal 2003; Baldry
\etal 2006).  Recent observational evidence suggests that growth of the red sequence since $z \sim
1$ is dominated by increasing numbers of low-luminosity ($\la L^*$) red galaxies with cosmic time at
the expense of the low-luminosity star-forming galaxy population (e.g. Bell \etal 2004; Bundy \etal
2006; Brown \etal 2007; Faber \etal 2007).  Hence, our understanding of growth and evolution in the
red sequence is closely related to our knowledge of processes responsible for quenching star
formation in blue-sequence galaxies, enabling their subsequent evolution onto the red sequence. 

The dichotomy of observed galaxy colours carries with it an increasingly well-described
environmental dependence whereby galaxies inhabiting high-density regions are on average redder,
older, more concentrated and more massive than galaxies found in the field (e.g.  Oemler 1974;
Dressler 1980; Goto \etal 2003; Blanton 2005; Blanton \& Berlind 2007; O'Mill, Padilla \& Lambas
2008; Cooper \etal 2008).  This strong correlation of red-sequence galaxies with local galaxy
density has led to a significant body of work devoted to exploring the environmental factors
responsible for creating these predominantly red populations from the blue, star-forming population
of field galaxies.  Of the possible mechanisms for suppressing star formation, ram-pressure stripping
(e.g. Gunn \& Gott 1972; Hester 2006), mergers (e.g. Toomre \& Toomre 1972; Negroponte \& White
1983; Somerville \& Primack 1999; McIntosh \etal 2008), strangulation (e.g. Larson, Tinsley \&
Caldwell 1980; Balogh \& Morris 2000), tidal interactions and harassment (e.g. Farouki \& Shapiro
1981; Moore \etal 1996) are the most frequently invoked to account for the increasing number density
of red galaxies.  Identifying {\it which} of these possible processes represents the dominant
evolutionary path, however, remains an active field of research.

The Sloan Digital Sky Survey (SDSS) has proved fundamental in exploring the environmental dependence
of galaxy properties, and has been used to show that the transformation of galaxies from blue to red
takes place over relatively long timescales, $\ga 1$\,Gyr, through the slow exhaustion of gas
available for star formation, rather than rapid, violent removal (e.g. Kauffmann \etal 2004; van den
Bosch \etal 2008).  The dominance of so called ``strangulation'' (Balogh \& Morris 2000) precludes
cluster-cores as dominant sites for blue galaxy transformation as disruption processes in
high-density regions, e.g. ram-pressure stripping and harassment, act over relatively short
timescales (e.g. Abadi, Moore \& Bower 1999; Quilis, Moore \& Bower 2000).  This supports earlier
work from the 2dF Galaxy Redshift Survey (2dFGRS) which found that the fraction of star forming
galaxies is lower with respect to the field even at the relatively low surface densities of loose
groups, $\sim$1-2 galaxies/Mpc$^2$, i.e. well outside dense cluster cores (e.g. Lewis \etal 2002;
Poggianti \etal 2006).

While the majority of our understanding of the variation of galaxy properties with environment
hinges on large photometric surveys, an alternative approach is the use of spectral absorption
features as detailed tracers of galaxy stellar populations.  In many ways spectral analyses provide
a vastly increased level of detail for determining paths of galaxy formation and evolution relative
to photometry, but they are often hindered by smaller sample sizes and relatively high
observational expense.  Nevertheless, absorption-line studies have met with great success in tracing
the stellar populations of individual galaxies (e.g. Proctor \etal 2004; Thomas \etal 2005;
S\'anchez-Bl\'azquez \etal 2006; Peletier \etal 2007; Spolaor \etal 2008), clusters (e.g. Smith,
Lucey \& Hudson 2007; Smith \etal 2008a,b; Trager \etal 2008) and large survey samples (e.g.
Gallazzi \etal 2006; Jimenez \etal 2007; Proctor \etal 2008).  

To date, spectroscopic studies with a particular focus on environment have primarily made use of
cluster galaxy populations.  Caldwell, Rose \& Concannon (2003) used a sample of nearby early-type
galaxies in the Virgo Cluster, as well as a field-galaxy sample, to probe galaxy stellar populations
to low masses (stellar velocity dispersions, $\sigma, < 100$\,\kms).  Caldwell \etal (2003) found a
greater intrinsic scatter in the properties of low-mass galaxies, but made no particular effort to
examine the spatial distribution of galaxy populations within the Virgo Cluster.  In similar work,
Smith \etal (2007) and Smith \etal (2008a,b) have examined the stellar populations of dwarf galaxies
in the Coma Cluster and Shapley supercluster, in both instances finding that the scaling relations
of low-mass populations are generally consistent with the relations of higher-mass E and S0
galaxies, although with the same increased scatter noted by Caldwell \etal (2003).  In the case of
Coma, Smith \etal (2008a) find a radial dependence of both age and metallicity, where galaxies at
larger projected radii are both younger and higher metallicity, suggestive of recently quenched star
formation among galaxies entering the Coma cluster outskirts and consistent with earlier spectral
studies of Coma cluster galaxies (Caldwell \etal 1993; Mobasher \etal 2001; Carter \etal 2002).  

These studies, however, only serve to probe rare high-density cluster environments, neglecting the
much more common group environment.  In an attempt to fill this gap, here we undertake a detailed
spectroscopic study of the galaxy population in the NGC\,5044 group with the aim of describing
galaxy evolutionary histories and reconstructing the formation history of the group as a whole.  In
a previous paper (Mendel \etal 2008; hereafter Paper I), we used new, deep spectroscopic
observations in conjunction with archival velocity information to more than triple the number of
velocity-confirmed group members.  This greatly improved sample has allowed us to characterise the
dynamical state of the group, which we find to be relaxed from a combination of X-ray and dynamical
indicators.  Here, we turn our focus to analysing the stellar populations of galaxies in this
redefined group with a focus on establishing the details of galaxy evolution in the context of the
group environment.    

The structure of this paper is as follows: Section \ref{obs} describes the spectroscopic data used
from both new AAOmega observations and the 6dF Galaxy Survey.  Here we also describe the method used
to derive kinematic properties. In Section \ref{index_measure} we discuss the measurement of Lick
index absorption features and their interpretation in terms of age, metallicity and $\alpha$-element
abundance (Section \ref{fitting}).  In subsequent sections we discuss the stellar population
properties of NGC\,5044 group galaxies both as an independent galaxy sample (Section
\ref{gal_prop}), and in the context of their place in the group as a whole (Section
\ref{group_params}).  In Section \ref{emm_line} we describe the distribution of emission-line
properties for NGC\,5044 group galaxies.  In Section \ref{conclusions} we discuss our stellar
population findings in relation to the dynamical description of the group presented in Paper I, as
well as their relevance to the formation and evolution of the NGC\,5044 group and galaxies therein.

Throughout this paper we assume $H_0$ = 70\kms\,Mpc$^{-1}$ where relevant.  We adopt a distance
modulus for the NGC\,5044 group of $(m-M)_0=32.31$ (28.99\,Mpc) measured using surface brightness
fluctuations (Tonry \etal 2001) with corrections applied to adjust for the improved Cepheid distance
measurements of Jensen \etal (2003).  Where specified, correlations have been measured using the
Kendall rank-order correlation test with significance quoted from the two-tailed p-value.

\section{Data} 
\label{obs}

Here we use the NGC\,5044 group sample defined in Paper I, which consists of 111
spectroscopically-confirmed group members compiled using newly obtained redshifts from AAOmega
supplemented using the 6dF Galaxy Survey Data Release 2 (6dFGS DR2\footnote{The most recent data
release available at the time of writing.}; Jones \etal 2005), HI observations (Kilborn \etal in
preparation) and the Nasa/IPAC Extragalactic Database (NED) sources.  Of these 111 galaxies, 95
have spectra available from AAOmega and 6dFGS observations which can be used to measure absorption
indices (71 from our new AAOmega observations and 24 from 6dFGS DR2).  These spectral data are
summarised below.

\subsection{AAOmega spectroscopy}
\label{aaospec}

New observations of NGC\,5044 group galaxies were carried out using the AAOmega multi-object
spectrograph on the 3.9m Anglo-Australian Telescope (AAT) in Siding Spring, Australia.  Target
galaxies were taken from the list of potential NGC\,5044 group members of Ferguson \& Sandage
(1990).  Medium resolution 580V and 385R gratings were used, yielding FWHM dispersions of
3.5\,\AA~and 5.3\,\AA~for blue ($\lambda\lambda$3700-5700\AA) and red ($\lambda\lambda$5700-8800\AA)
spectra respectively.

Galaxies were separated into high-- and low--luminosity subsamples using an apparent magnitude cut
of $B=15$ and observed separately in order to limit the effects of scattered light from the
brightest sources.  Observations were carried out 30 minutes at a time with a total of 3 to 4
observations per plate configuration.  In addition to fibres assigned to galaxies, in each
configuration an additional 30 to 35 ``empty'' fibres were assigned to be used for sky subtraction.
There are several relatively wide defects in the blue arm CCD of AAOmega, and to limit the effects
of these on our final coadded science spectra we offset the central wavelength on each night of
observations by 100\,\AA, ensuring that usable data were obtained from at least two out of three
nights at every wavelength.  Our total integration time for high- and low-luminosity galaxies was 4
and 17 hours respectively.  

In addition to our primary (galaxy) targets we have observed a set of calibration stars.  These
include stars common to the Lick stellar library (see $\S$\ref{lick_errors}) and a set of
spectrophotometric calibrators which we can use to roughly flux calibrate the AAOmega data.

Data were reduced using a combination of {\sc iraf} routines and the {\sc 2dfdr} reduction pipeline
supplied and maintained by the Anglo-Australian Observatory.  Prior to reduction with {\sc 2dfdr} an
average bias was subtracted from all frames and bad pixel columns were identified and repaired.  We
then used {\sc 2dfdr} to reject cosmic rays, identify fibre apertures, flat-field, throughput
calibrate and sky subtract the data.  The final coaddition of science frames was carried out after
excluding any single exposures with very low flux relative to the other frames. 

In this work, we consider only those galaxies identified as group members in Paper I, and refer the
reader to that work for the details of how group membership was defined.  We exclude from subsequent
analyses any galaxies with a signal-to-noise (S/N) of less than 12, as below this we consider
stellar population measurements to be highly unreliable.  This cut excludes 19 galaxies, leaving a
total of 52 from the original AAOmega sample described in Paper I.

\subsection{6dFGS Spectral Data}
\label{6dfspec}

The 6dFGS covers 17046 deg$^2$ of the Southern sky using the Six-Degree Field (6dF) multi-fibre
spectrograph on the UK Schmidt Telescope.  The survey targets are selected from the 2MASS Extended
Source Catalogue (2MASS XSC; Jarrett \etal 2000), 2MASS and SuperCOSMOS to include all galaxies
brighter than $K_{\mathrm{tot}}=12.75$\,mag with wavelength coverage from
$\lambda\lambda$4000-8400\AA.  In Paper I we include 16 unique 6dFGS galaxies and 8 galaxies that
overlap with our AAOmega sample, all of which have spectra available.  However, applying a similar
cut in S/N as the AAOmega data above excludes an additional 9 of the 6dFGS galaxies, leaving a total
of 15 for stellar populations analysis. 

The combined AAOmega and 6dFGS sample then consists of 67 confirmed group-member galaxies with S/N
$> 12$, which we use for the remainder of this analysis.

\subsection{Recession velocity and velocity dispersion measurements}
\label{rv_sig}

Recession velocities and velocity dispersions were fit simultaneously using the penalised
pixel-fitting (pPXF) software of Cappellari \& Emsellem (2004).  Pixel fitting methods are
notoriously sensitive to the templates used, so it is important to minimise the effects of so-called
template mismatch on kinematic measurements.  The pPXF code is an extension of the Gauss-Hermite
fitting method of van der Marel \& Franx (1993) which allows for fitting to be carried out using a
linear combination of template spectra, all but eliminating template-mismatch-induced errors.

As our input templates we use 50 stellar spectra from the MILES standard star library
(S\'anchez-Bl\'azquez \etal 2006) spanning a broad range of spectral types.  The particular choice
of stars used to construct templates has only a minimal effect on the results derived using pPXF as
long as they span a similar range of spectral types.  We have investigated this briefly by fitting a
sub-sample of our galaxy data with sets of 50 randomly chosen stars from the MILES library and find
typical deviations of $\sim$4-5\,\kms in velocity dispersion.  

In order for pPXF to provide meaningful velocity dispersion outputs we need to account for the
difference in instrumental dispersion between our AAOmega spectra and the MILES stellar library.
Overlap between the Lick and MILES stellar libraries is significant, so we have used the Lick
standard stars observed throughout our AAOmega observing run to effectively measure the wavelength
dependent broadening offset between AAOmega and the MILES library.  In total we use 5 Lick/MILES
stars to measure the median broadening difference.  We find that scatter in our broadening
measurements is significant for wavelengths below 4700\,\AA, but above this wavelength the
dispersion offset is relatively stable at $\sim$57.7\kms (2.3\,\AA~FWHM).  Adding in quadrature the
spectral resolution of the MILES library, 2.3\,\AA, gives an average spectral resolution for AAOmega
between 4700\,\AA~and 5400\,\AA~of 3.2\,\AA~FWHM.  This derived resolution is consistent with that
found by Smith \etal (2007) for spectra taken using the 580V grating on AAOmega.

Standard star observations for the 16 6dFGS galaxies that we define as group members are
unavailable, but we can measure the broadening offset between AAOmega and 6dF using the 8
galaxies present in both samples.  In the $\lambda\lambda$4700-5400\,\AA~wavelength range we find a
mean offset between AAOmega and 6dFGS spectra of 103.6\kms (4.1\,\AA~FWHM).  Adding in quadrature
the previously measured offset between AAOmega and the MILES library in this spectral region we
adopt a total broadening of 118.6\kms (4.7\,\AA~FWHM). The difference in aperture size between the
6dF and AAOmega fibres (6.7$^{\prime\prime}$ vs. 2.1$^{\prime\prime}$) will result in an additional
broadening difference between the two galaxy spectra, independent of instrument resolution, of order
5 percent (J\o rgensen, Franx \& Kj\ae rgaard 1995).  We do not take this aperture effect into
account when calculating our broadening as the statistical nature of the J\o rgensen \etal (1995)
correction will likely only serve to introduce scatter in our relatively small number of galaxies,
particularly given our broad range of galaxy types.

Prior to measuring kinematics using pPXF, MILES template spectra are smoothed to the AAOmega and
6dFGS resolution using the measured broadening offsets.  The pPXF code allows for a penalty to be
applied to the higher order terms of the Gauss-Hermite polynomials and here we adopt a moderately
high bias of 0.8 throughout fitting in order to limit spurious results from undersampled or low S/N
data.  Random errors on our kinematic measurements are then calculated using a series of Monte Carlo
simulations spanning a range of velocity dispersions and S/N.  The scatter in our determination of
instrumental broadening discussed above adds an additional, systematic uncertainty to our velocity
dispersion measurements of 22\kms and 41\kms for AAOmega and 6dFGS galaxies respectively, which we
include in quadrature.

\subsection{Emission-line measurements}
\label{emission_measure}

We identify emission-line galaxies through examination of residuals to the best fit pPXF templates,
and for 24 galaxies in our sample we find evidence for emission in some combination of H$\beta$,
[OIII]$_{\lambda 4959}$, [OIII]$_{\lambda 5007}$, [NI]$_{\lambda 5198}$ and [NI]$_{\lambda 5200}$.
While we can estimate emission-line fluxes from the residuals of our template fits, the underlying
absorption in these regions is generally poorly constrained as emission contaminated lines are
masked in the template fitting process.  
 
A far more robust method is to fit for absorption- and emission-line kinematics simultaneously,
which we carry out using the {\sc gandalf} software of Sarzi \etal (2006).  This is an extension of
the pPXF pixel fitting routine described in $\S$\ref{rv_sig} which fits simultaneously a set of
absorption and emission templates, eliminating errors introduced by fitting emission line residuals
separately.  Fits to emission lines are conducted using a set of independent Gaussian profiles whose
kinematics can be varied as required.  Here we treat the kinematics of nebular and Balmer-line
emission separately, tying each to the kinematics of the dominant features in appropriate regions of
our spectra (H$\beta$ or H$\alpha$ and [OIII]$_{\lambda 5007}$ or [NII]$_{\lambda 6583}$ for Balmer
and nebular lines respectively).  Our choice to fit separately the kinematics of Balmer and nebular
emission lines has a negligible effect on our results as we find no galaxies with a significant
kinematic difference between the two (where both are detected).  There is generally good agreement
between kinematics measured using pPXF and {\sc gandalf}; the largest offsets are observed in
velocity dispersion, and are only significant in galaxies with strong emission lines and relatively
weak continuum flux.
 
{\sc gandalf} also allows for spectra to be cleaned of detected emission.  Here we follow Sarzi
\etal (2006) and adopt an amplitude-to-noise (A/N) threshold of 4, where noise is defined as the
scatter about the best-fit absorption template.  As both [NI] and H$\beta$ emission strongly affect
our Lick index measurements the removal of these lines results in a significant improvement in the
quality of stellar population fits for emission-line galaxies (see discussion in Section
\ref{fitting}).  The approximate detection threshold of emission lines in our data can be calculated
as a function of S/N using our adopted A/N and typical line broadening.  If we assume a typical
intrinsic line width of $\sim$30--50\kms, then our sensitivity is $\sim$0.6\,\AA~for AAOmega data
and $\sim$0.9\,\AA~for 6dFGS data at the median S/N of our galaxy data.

We use the H$\alpha$/H$\beta$ Balmer decrement to calculate the extinction correction for our
emission-line fluxes where applicable using the $R_V=3.1$ reddening curve of Cardelli, Clayton \&
Mathis (1989) and an intrinsic H$\alpha$/H$\beta$ line ratio of 2.85 (Osterbrock 1989).  In galaxies
with very low extinction the errors on our line-flux measurements can sometimes result in a negative
extinction estimate.  As this is clearly non-physical we adopt  $E(B-V)\leq0.01$ as an upper limit
of extinction and assign this to galaxies with low or spurious (i.e. negative) extinction
measurements.

\section{Lick index measurements and calibration to the Lick/IDS system}
\label{index_measure}

Here we use Lick absorption features (Burstein \etal 1984; Trager \etal 1998) to characterise the
stellar populations of our galaxies.  Lick indices are defined based on their particular sensitivity
to either age or metallicity, and so are useful for describing these parameters in integrated
spectra.  Lick line-strength measurements were carried out using the index definitions of Trager
\etal (1998), supplemented with higher-order Balmer line definitions from Worthey \& Ottaviani
(1997).  We omit from our Lick index measurements Fe5782, NaD, TiO$_1$ and TiO$_2$ as these fall in
the wavelength range of the AAOmega dichroic where sensitivity is decreasing rapidly.

\subsection{Velocity dispersion corrections}
\label{vdisp_corr}

In order to properly compare measured indices with SSP models it is important that the total
broadening of any given absorption feature matches the wavelength-dependent broadening of the
Lick/IDS system.  Galaxies with combined instrumental and velocity-dispersion broadening less than
the target Lick resolution are broadened as necessary using a wavelength-dependent Gaussian kernel
to match the description of the Lick/IDS resolution given by Worthey \& Ottaviani (1997). 

In cases where the combined instrument and velocity dispersion broadening is greater than the
required Lick/IDS system resolution we calculate line-width corrections based on iteratively
broadened template stars.  These line-width adjustments are then applied to correct measured indices
back to the desired Lick/IDS resolution.  As a test of this method, Proctor \& Sansom (2002) compare
stellar and galaxy spectra broadened using this technique, finding good agreement between the two.
For further details see Proctor \& Sansom (2002).

\subsection{Lick/IDS system calibration}
\label{lick_errors}

In order to calibrate index measurements to the Lick/IDS system, we have observed several stars from
the Lick stellar library at the beginning of each night.  Index measurements for AAOmega data are
then adjusted based on a comparison of published Lick standard star indices and our stellar
observations.  The mean offsets found for each index are listed in Table \ref{aao_offsets}, along
with their associated r.m.s. error. 

We are unable to calibrate 6dFGS data in the same way as AAOmega data, and so we generate two sets
of indices for 6dFGS galaxies: the first using the corrections provided in Table \ref{aao_offsets},
and the second with no correction.  We refer the reader to $\S$\ref{aao_6df} for a comparison of the
properties derived using these two sets of indices for overlapping 6dFGS and AAOmega galaxies.

\begin{table}
\centering
\caption{Lick index corrections for AAOmega data and their associated errors.} 
\begin{tabular}{lrc}
\hline
Index & Offset & $\sigma_{rms}$ \\
\hline
H$\delta_A$ & 0.148 & 0.426  \\
H$\delta_F$ & 0.053 & 0.208  \\
CN$_1$      & 0.018 & 0.008  \\
CN$_2$      & 0.024 & 0.010  \\
Ca4227      & 0.153 &  0.098  \\
G4300       & --0.029 & 0.110  \\
H$\gamma_A$ & 0.244 & 0.255  \\
H$\gamma_F$ & 0.099 & 0.118  \\
Fe4383      & 0.362 & 0.093  \\
Ca4455      & 0.386 & 0.049  \\
Fe4531      & 0.025 & 0.107  \\
C4668       & --0.569 & 0.197  \\
H$\beta$    & --0.133 & 0.039  \\
Fe5015      & 0.222 & 0.074  \\
Mg$_1$      & 0.028 & 0.002  \\
Mg$_2$      & 0.036 & 0.002  \\
Mg{\it b}   & 0.006 & 0.051  \\
Fe5270      & --0.096 & 0.125  \\
Fe5335      & --0.004 & 0.055  \\
Fe5406      & --0.014 & 0.039  \\
Fe5709      & 0.183 & 0.090  \\
\hline
\end{tabular}
\label{aao_offsets}
\end{table}

\subsection{Index errors}
\label{ind_errors}

The fitting technique that we adopt for these data involves the simultaneous $\chi^2$ minimisation
for a large number of Lick indices ($\sim$20).  In order for these fits to be reliable, an accurate
representation of index errors is crucial, but notoriously difficult to obtain for spectral
observations.   

Proctor \etal (2008; hereafter P+08) have recently used spectra from the 6dFGS DR1 (Jones \etal
2004) to measure ages and metallicities in a large sample of galaxies using Lick indices.  In
estimating their index errors, P+08 have used duplicate observations of the same galaxy (from
overlapping survey regions) to calculate the r.m.s. scatter in their Lick index measurements at a
given S/N.  The r.m.s. value was then fit as a function of S/N using the form
$\mathrm{r.m.s.}=a/(\mathrm{S/N}+b)$, where $a$ and $b$ are constants, which was then used to
calculate representative errors for the rest of the galaxy sample based on their S/N.  This method
not only encapsulates redshift and velocity dispersion errors, but also errors resulting from sky
subtraction, fibre flat-fielding and throughput calibration that are nearly impossible to quantify
in individual fibre observations. 

\begin{figure}
\centering
\includegraphics[scale=0.35,angle=-90]{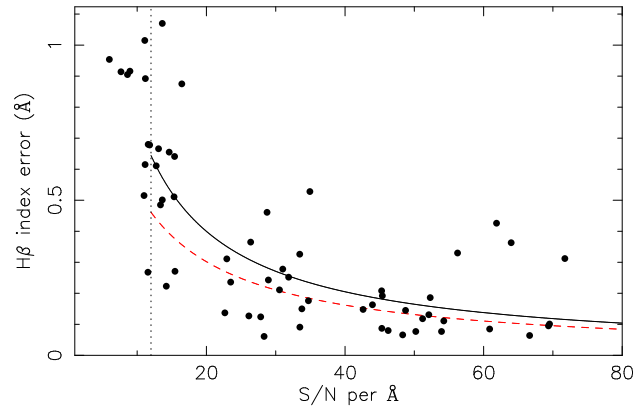}
\caption{H$\beta$ index error vs. signal-to-noise for galaxies in our spectroscopic sample.  The
{\it solid} line shows the best fit to our measured index errors as a function of S/N (see
$\S$\ref{ind_errors}), while the {\it dashed} line shows a similar fit for 6dFGS DR1 data derived by
P+08.  The {\it dotted}, vertical line shows our adopted S/N cut of 12.} 
\label{hb_snr}
\end{figure}

Here we adopt a similar strategy to P+08 and exploit the composite nature of our final spectra by
re-measuring indices on individual observations, of which there are 12 and 34 observations each for
bright and faint galaxies respectively.  We then adopt the $\sigma_{rms}$ for these repeat
measurements as an accurate representation of the stochastic errors in our sample (i.e. Poisson
noise, sky-subtraction etc).  In Fig. \ref{hb_snr} we show how the H$\beta$ index error derived in
this manner varies as a function of S/N for the galaxies in our sample, and for comparison we also
show the line describing index errors as derived by P+08 for 6dFGS DR1 data.  Error curves for all
indices are shown in Appendix \ref{index_errors}.

For our AAOmega data the total adopted index error is a combination of this random error, the
$\sigma_{\mathrm{r.m.s.}}$ associated with conversion to the Lick/IDS system ($\S$\ref{lick_errors})
and the index broadening due to errors on velocity dispersion and recession velocity measurements
($\S$\ref{rv_sig}).  For 6dFGS data we adopt the recession velocity and velocity dispersion errors
as discussed in $\S$\ref{rv_sig}, as well as the index errors as a function of S/N derived by P+08. 

\section{Stellar population models and fitting}
\label{fitting}

Having compiled relevant Lick index measurements and their corresponding error, we now turn to the
task of interpreting these measurements in terms of the stellar population parameters they describe.
Here we use the Single Stellar Population (SSP) models of Thomas, Maraston \& Bender (2003) with
account for variable abundance ratios as calculated by Korn, Maraston \& Thomas (2005; hereafter we
refer to the Thomas \etal 2003 and Korn \etal 2005 models collectively as TMB03).  These models
cover the metallicity range $-$2.25$\leq$[Z/H]$\leq$0.65 with ages from 1 to 15\,Gyrs.  The variable
abundance ratios included by TMB03 consider N, O, Mg, Ca, Na, Si and Ti as enhanced elements and,
while these fall outside a strict definition of $\alpha$-elements, we will hereafter refer to
enhancement measures from the TMB03 models as $\alpha$-element abundances or [$\alpha$/Fe].

It is well known that spectra are degenerate with respect to age and metallicity in so much as old,
metal-rich and young, metal-poor stellar populations are remarkably similar.  In order to minimise
the influence of this degeneracy on our final measurements, we adopt the $\chi^2$ minimisation
technique of Proctor \etal (2004), simultaneously fitting as many Lick indices as possible.  This
method has been shown to recover reliable ages, metallicities and $\alpha$-element abundances in
both galaxies (e.g.  Proctor \& Sansom 2002; Brough \etal 2007; Proctor \etal 2008) and globular
clusters (Beasley \etal 2004; Pierce \etal 2006; Mendel \etal 2007).  

Our goal is to obtain a stable fit between our galaxy data and the TMB03 models, and to accomplish
this we adopt an iterative approach to our fitting.  As a first step we perform a fit using as many
indices as our data allow, generally between 18 and 21.  In subsequent iterations indices deviant
from the best-fit solution at the 5, 4 and 3 $\sigma$ levels are clipped.  A final, manual
inspection of the fit is then carried out to ensure that stable fits have been attained and that
indices have not been over- or under-clipped as a result of poor error estimates.  

\begin{figure}
\centering
\includegraphics[scale=0.34,angle=-90]{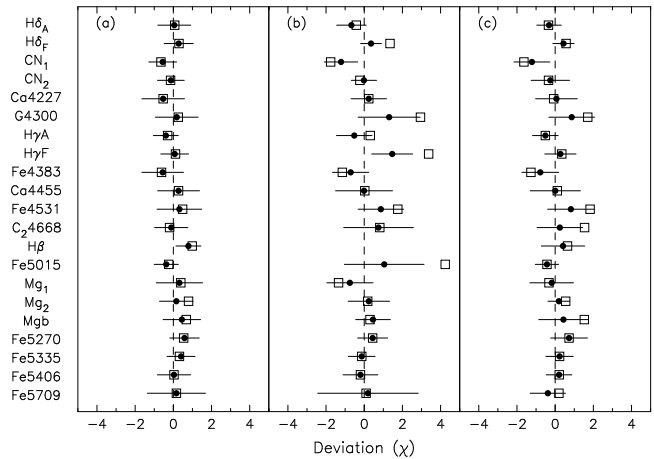}
\caption{Mean deviation of indices from the best fit TMB03 SSP model.  Open squares show the mean
deviation for all indices in each galaxy, while filled circles show the mean deviation only for
indices included in the fits. (a) Galaxies with no detectable emission, (b) galaxies with visible
emission in their spectra and (c) emission galaxies with spectra cleaned using the {\sc gandalf}
routine (see $\S$\ref{emission_measure}).} 
\label{chis}
\end{figure}

\begin{figure}
\centering
\includegraphics[scale=0.34,angle=-90]{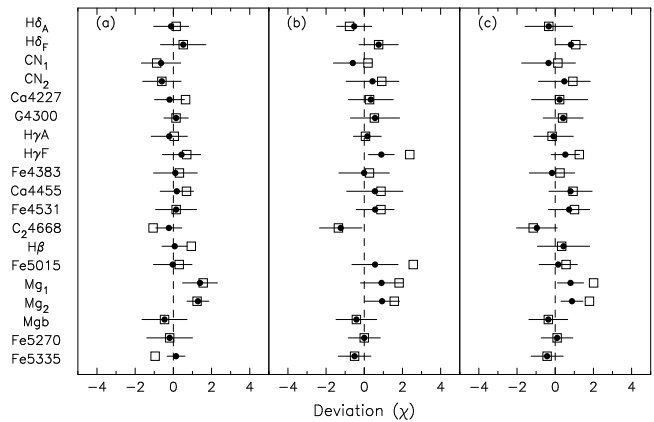}
\caption{Same as Fig. \ref{chis}, but for 6dFGS data.  Data shown are without correction to the
Lick/IDS system (see $\S$\ref{aao_6df}).} 
\label{6df_chis}
\end{figure}

Figs. \ref{chis} and \ref{6df_chis} show the mean deviation of indices from the best fit model for
AAOmega and 6dFGS data respectively.  Galaxies free of emission are generally fit very well by the
TMB03 models and we clip at most 2 indices from any given galaxy.  Emission-line galaxies are
significantly more difficult to fit, however the emission-cleaned spectra output from {\sc gandalf}
result in a large improvement to the overall quality of fits for AAOmega data (Fig \ref{chis}b,c)
and slight improvement for 6dFGS data (Fig.  \ref{6df_chis}b,c).

Final errors on our stellar population measurements are estimated using a Monte Carlo method which
re-samples the best-fit model convolved with our observed index errors.  These error estimates
represent our best attempt to quantify the random error contribution to our measurements, however
there is an additional systematic error present in the models themselves which we do not account for
(see, e.g., Conroy, Gunn \& White 2008).

The 6dFGS and AAOmega spectra used in this work can only be used to derive {\it luminosity-weighted}
ages, metallicities and $\alpha$-element abundances.  This is important to any subsequent analyses
as these luminosity-weighted values are particularly affected by small centralised bursts of star
formation and may not be representative of the global galaxy properties.  In addition, because of
aperture size differences between AAOmega and 6dF (2.1$^{\prime\prime}$ vs. 6.7$^{\prime\prime}$;
0.3\,kpc vs. 0.9\,kpc at the distance of NGC\,5044) we expect some variation in the derived stellar
population parameters due to both sampling differences and the presence of radial gradients,
predominantly in metallicity.  Final measurements of stellar population parameters (ages,
metallicities etc.) and their associated errors for the 67 galaxies in our spectral sample are
included in Appendix \ref{sp_app}. 

\subsection{Comparison of AAOmega and 6dFGS galaxies: agreement and aperture effects}
\label{aao_6df}

\begin{figure*}
\centering
\includegraphics[scale=0.63,angle=-90]{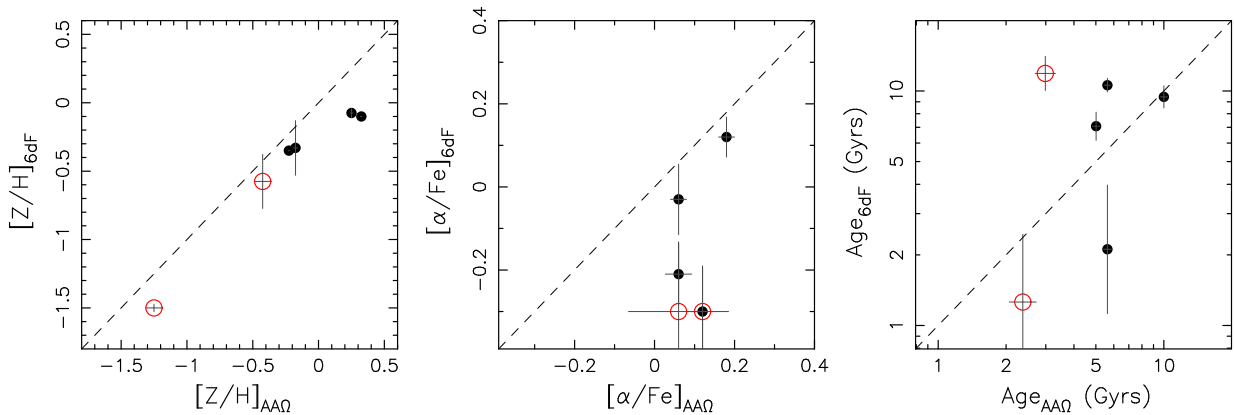}
\caption{Comparison of metallicity ([Z/H]), $\alpha$-element abundance and age for the 6 overlapping
6dFGS and AAOmega galaxies with S/N $>$ 12.  Open (red) and filled (black) circles represent
galaxies with and without emission lines.  {\it Dashed} lines in each panel represent equality.
Parameters for 6dFGS data are shown with no applied correction to the Lick/IDS system (see
$\S$\ref{lick_errors}).} 
\label{sdf_aao_comp}
\end{figure*}

In Fig. \ref{sdf_aao_comp} we show the comparison of stellar population parameters derived for
overlapping AAOmega and 6dFGS galaxies.  In this comparison we have used indices measured from 6dFGS
data with no Lick/IDS system correction applied.  Using indices adjusted using the AAOmega Lick/IDS
system correction gives similar results, but with an increase in scatter.  We therefore use
uncorrected 6dFGS indices for the remainder of this work.

Of particular interest here is the comparison between galaxy properties sampled using the different
fibre apertures of AAOmega and 6dF.  Perhaps the most obvious trend is observed in metallicity,
which is also the most robust of our stellar population measurements.  6dFGS galaxies show a
systematic offset towards lower metallicities of $\sim$0.1\,dex, consistent with observations that
the majority of galaxies possess negative gradients in metallicity (i.e. galaxies are more metal-poor
at larger radii, e.g. S\'anchez-Bl\'azquez \etal 2007; Brough \etal 2007).  

Comparisons of $\alpha$-element abundances and ages are considerably more scattered than
metallicity, indicative of the increased difficulty in their measurement.  Broadly, we expect the
larger fibre aperture of 6dF to yield older ages, particularly for galaxies with recent or currently
ongoing star formation as these bursts are (generally) centrally concentrated.  We find this to be
the case for half of the 6dFGS galaxies, but the other half scatter to younger ages with
considerable uncertainty.  As these galaxies are just above our adopted S/N cut, we consider these
measurements to be highly uncertain rather than representative of a real trend in the data.  We
expect the variation in aperture size between AAOmega and 6dF to have little systematic effect on
the measured $\alpha$-element abundances, as observed gradients in [$\alpha$/Fe] are both weak and
variable.  In Fig. \ref{sdf_aao_comp}, $\alpha$-element measurements using AAOmega data are
consistently higher than those from the 6dFGS data.  Again, this is most likely a S/N effect as the
small dynamic range of $\alpha$-element abundances makes their determination difficult in low S/N
data. 

Motivated by the comparisons in Fig. \ref{sdf_aao_comp} and the discussion above, in forthcoming
sections we consider galaxy metallicities to be robust for both the 6dFGS and AAOmega samples.  Age
and $\alpha$-element abundances for 6dFGS galaxies can be used to give a rough indicator of stellar
population trends at larger group-centric radii (see $\S$\ref{pop_radius} and Paper I), but we will
refrain from using them to make any specific judgements due to the large uncertainty in their
relationship to AAOmega data.

\section{Galaxy properties}
\label{gal_prop}

\subsection{Stellar population parameters versus stellar mass}
\label{sp_params}

\begin{figure}
\centering
\includegraphics[scale=0.65,angle=-90]{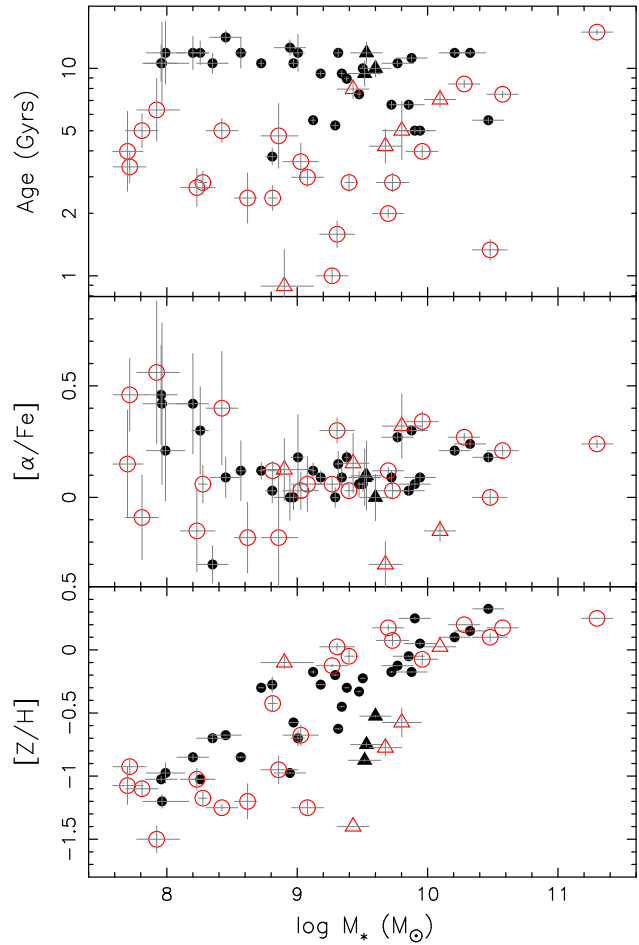}
\caption{Relationship between stellar population properties and stellar mass (see
$\S$\ref{sp_params}).  Filled (black) and open (red) symbols represent passive and emission-line
galaxies (see $\S$\ref{emission_measure}), with circles and triangles delineating galaxies measured
using AAOmega and 6dFGS spectra, respectively.  Errors bars represent 1-sigma deviations estimated
using the Monte Carlo technique described in Section \ref{fitting}.} 
\label{mass_sp}
\end{figure}

Before attempting to place our observed galaxy population in the context of the NGC\,5044 group
environment, we first examine the properties of galaxies independent of their particular place in
the group.  In Fig. \ref{mass_sp} we show how our measured ages, metallicities and $\alpha$-element
abundance vary with stellar mass ($M_*$).  Stellar masses have been computed using GALEXEV stellar
population synthesis models of Bruzual \& Charlot (2003; hereafter BC03).  Here we use BC03 models
constructed using the ``Padova 1994'' isochrones (see BC03 and references therein) and stellar initial
mass function (IMF) of Chabrier (2003).  Using our derived central ages and metallicities, we then
extract the appropriate {\it B}-band mass-to-light ratio (M/L) from the BC03 models. 

Of the three stellar population parameters measured and shown in Fig.  \ref{mass_sp}, metallicity
shows the strongest correlation with mass, spanning more than two decades with a scatter of
$\sim$0.26\,dex.  The general tightness of the mass--metallicity relation as determined from both
emission- and absorption-line analyses is well known from large samples such as the SDSS (e.g.
Tremonti \etal 2004; Gallazzi \etal 2006), however our data probe this relation an order of
magnitude lower in both mass and metallicity.  We see evidence for increased scatter among
emission-line galaxies ($\sim0.39$\,dex for emission-line galaxies vs. $\sim0.23$\,dex for passive
galaxies), which could be indicative of ``contamination'' from recently formed stellar populations,
however the fact that metallicities for low-mass emission-line galaxies ($\log M_* \la 9$) appear
depressed relative to their passive counterparts suggest that contamination is not playing a
significant role, as recently formed populations should be metal-enriched.  Instead, the observed
low metallicities (as well as increased scatter) could be an effect of metal-enriched outflows in
these galaxies, driven either by starbursts or supernovae, which could prevent the incorporation of
metals into newly-formed stars.  We note, however, that measurements for emission-line systems are
also affected by the quality of our applied emission-line corrections, which would naturally lead to
an increased scatter in absorption-line measurements in these systems.

\begin{figure*}
\centering
\includegraphics[scale=0.60,angle=-90]{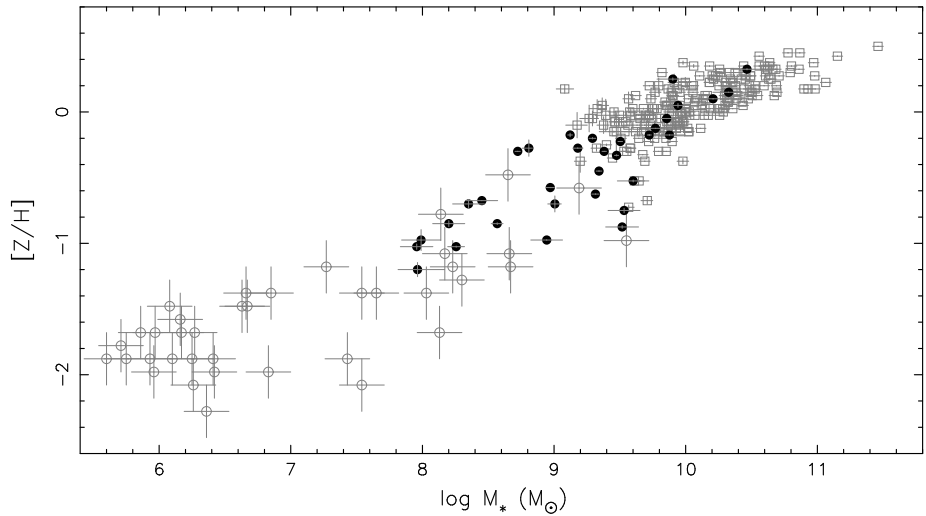}
\caption{Mass--metallicity relation for galaxies spanning a broad range of environments.  Filled
circles, open circles and open squares represent data from the NGC\,5044 group, Local Group and
Shapley supercluster respectively.  Errors for the Local Group data are taken as the average error
quoted in Woo \etal (2008) of 0.17 and 0.2\,dex for stellar mass and metallicity.  Errors on
NGC\,5044 group and Shapley supercluster metallicities have been estimated using the Monte Carlo
technique described in Section \ref{fitting}.} 
\label{all_mass_met}
\end{figure*}

An interesting question related to the mass--metallicity relation is the degree to which it depends
on environment.  While a detailed analysis of this dependence requires a comparison of galaxies at
fixed stellar mass in a range of environments, even a qualitative comparison of the mass-metallicity
relation in several different environments should allow identification of any gross systematic
deviations.  To this end, we have selected a comparative sample of galaxies, both at higher and
lower masses, in two other environments, the Local Group and the Shapley supercluster.  Data for
Local Group dwarf galaxies are taken from the compilation of Woo, Courteau \& Dekel (2008), who
provide stellar masses, derived using the colour-$M_*/L$ relation of Bell \& de Jong (2001), and
average [Z/H] from red giant branch stars.  Metallicities for a sample of galaxies in the Shapley
supercluster have been calculated using Lick line index measurements from Smith \etal (2007) and the
fitting methods described in Section \ref{fitting}.  Stellar masses for these galaxies have been
estimated as described above, using $B$-band magnitudes from Smith \etal (2007).  A comparison of
the mass--metallicity relation for these three samples is shown in Fig.  \ref{all_mass_met}. In this
figure, metallicity estimates for Shapley supercluster galaxies have had additional offset of
+0.15\,dex applied to account for the different physical aperture size between NGC\,5044 and Shapley
observations (0.3\,kpc vs. 1.9\,kpc respectively)\footnote{Assuming an average metallicity gradient
of -0.25\,dex per dex, e.g. S\'anchez-Bl\'azquez \etal (2007)}.  Data shown in Fig.
\ref{all_mass_met} form a remarkably uniform mass--metallicity sequence over roughly six orders of
magnitude in galaxy stellar mass, despite the widely varied environments sampled.  Furthermore,
apart from the obvious differences in stellar mass range, the data suggest that environment is
likely playing only a small role in establishing the mass--metallicity relation, consistent with
recent results from the SDSS (e.g.  Mouchine \etal 2007; van den Bosch \etal 2007). 

Interpretations of the mass-metallicity relation generally focus around decreasing star-formation
efficiency at low masses, which can suitably explain the trends observed in galaxies (e.g.  Tremonti
\etal 2004; Savaglio \etal 2005; Gallazzi \etal 2006) and is reproduced in hydrodynamic simulations
including feedback mechanisms (e.g. Brooks \etal 2007).  In this interpretation of the
mass-metallicity relation we expect to see low-mass galaxies trend towards lower $\alpha$-element
abundances, as the relationship of $\alpha$-element abundance to supernovae timescales, namely SNII
versus SNIa contributions, provides leverage in separating between rapid, high-efficiency star
formation and on-going, low-efficiency star formation (e.g. Terlevich \& Forbes 2002; Thomas \etal
2005).

Examining the observed relation of [$\alpha$/Fe] with mass in Fig. \ref{mass_sp}, we find
considerable scatter, particularly in galaxies with emission.  The majority of this scatter results
from increasingly large index errors at low S/N (see Appendix \ref{index_errors}), but there is also
a systematic effect related to our fitting of low-mass, and hence low-metallicity, data.  The TMB03
model grids ``pinch'' together at low [Z/H], making the discrimination of [$\alpha$/Fe] increasingly
difficult (Mendel \etal 2007).  So, while we appear to observe significant scatter in the
[$\alpha$/Fe] measurements of low-mass galaxies, which would generally indicate significant
variation in star-formation histories, we are unable to draw any strong conclusions from these data.
If we restrict ourselves to galaxies with $\log M_* \ga 8.5\,M_{\mathrm{\odot}}$, then our data
exhibit a weak positive correlation of $\alpha$-element enhancement with mass at the $\sim$95
percent level, consistent with the interpretation of the mass--metallicity relation as indicative of
star-formation efficiency, discussed above.

In our age estimates we see a clear offset between emission-line and passive galaxies, unlike either
metallicity or $\alpha$-element abundance.  The mean stellar age of emission-line galaxies is
$\sim$3.6\,Gyrs, relative to $\sim$9.1\,Gyrs for passive galaxies.  We must, of course, interpret
these age measurements with caution as they are strictly luminosity-weighted measurements within our
fibre apertures.  Emission-line galaxies show a trend of younger central ages towards increasing
mass, significant at the $\sim$2$\sigma$ level, which we believe to be largely driven by aperture
effects and the centrally concentrated nature of star formation.  Analyses of central stellar
populations in conjunction with global photometric measurements allow one to constrain the relative
fraction of mass contained in central starbursts, which has been found to be of order 10 percent for
a range of galaxy masses (e.g. P+08).  If we consider the fact that the fraction of galaxy light
sampled by AAOmega's 2$^{\prime\prime}$ aperture varies by more than an order of magnitude from the
brightest to faintest sources in our sample, then the stellar population measurements for our most
massive galaxies are almost completely dominated by any central star formation, while in smaller
galaxies we sample a growing fraction of the underlying, old stellar populations. 

The passive sample of galaxies seems to separate into two sub-samples: one with relatively uniform,
old ages ($\ga9$\,Gyrs), and a second population with relatively young central ages.  This younger
sub-sample almost certainly represents galaxies which have more recently undergone bursts of star
formation and are now fading to older apparent ages.  One possible cause for this population could
be recent, gas-rich mergers which would serve to drive down central age measurements (e.g.
Kauffmann 1996).  In this scenario we would expect to see these star-formation bursts accompanied by
a decrease in $\alpha$-element abundance as new generations of stars form from increasingly
metal-enriched gas, however the $\alpha$-element abundances for these galaxies are consistent with
the bulk of the passive galaxy population.  

To summarise, the ages, metallicities and $\alpha$-element abundances of galaxies in the NGC\,5044
group are consistent with properties found in larger galaxy samples.  The tight mass--metallicity
relation and its interpretation as a sequence of star-formation efficiency is consistent with our
measurements of $\alpha$-element abundance, although the trend of elemental abundance with mass
appears relatively weak in these data. 

\subsection{The age--mass--metallicity relation}
\label{degeneracy}

The comparison of galaxy ages, metallicities and element abundances described in the previous
section demonstrates the close relationship between metallicity and stellar mass, but it should
also be considered that more complex relationships between multiple stellar population parameters
and mass may exist.  For example, in a study of local early-type galaxies Trager \etal (2000) found
that their elliptical galaxies were described by two separate two-dimensional relations in
four-dimensional space: a plane described by the linear combination of log\,$t$, log\,$\sigma$ and
[Z/H], and a relationship between [$\alpha$/Fe] and log\,$\sigma$ such that more massive galaxies
have higher values of $\alpha$-element enhancement.  Smith \etal (2008b) found a similar relation
between age, mass and metallicity for dwarf galaxies in the Coma cluster, noting in particular an
increase in scatter at low masses.

\begin{figure*}
\centering
\includegraphics[scale=0.63,angle=-90]{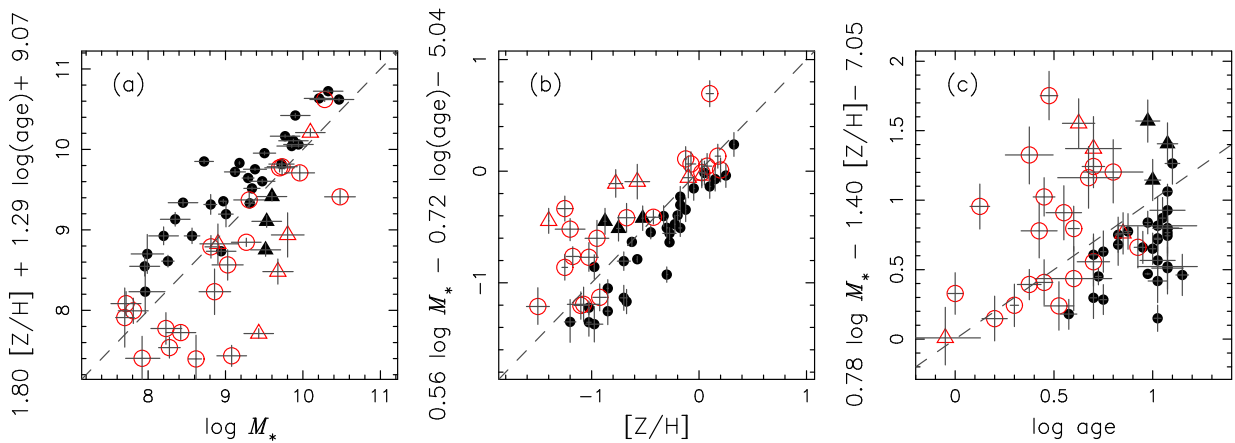}
\caption{Projections of the fitted plane to age, mass and metallicity.  Symbols are the same as Fig.
\ref{mass_sp}.  {\it Dashed} lines represent equality and are shown for reference.} 
\label{met_mass_age}
\end{figure*}

Motivated by recent comparisons suggesting a tight relationship between the properties of galaxies
at a fixed {\it stellar} mass (e.g. van den Bosch \etal 2008), we fit the plane described by our
data for age, mass and metallicity such that

\begin{equation}
\label{plane_eqn}
\log\,M_{\mathrm{*}}=\alpha\,\mathrm{[Z/H]} + \beta\log\,t + \gamma,
\end{equation}

\noindent using a least-squares fitting method and minimising residuals orthogonal to the plane.
The best-fitting parameters for these data result in a plane such that $\alpha = 1.83\pm0.46$,
$\beta = 1.41\pm0.79$ and $\gamma = 9.00\pm0.61$.  In Fig. \ref{met_mass_age} we show projections of
this best-fit plane in stellar mass, $M_{\mathrm{*}}$, metallicity, [Z/H], and age.  There is a
clear separation between passive and emission-line galaxies about the fitted plane, which is driven
primarily by the difference in mean age of the two populations and evident in Fig.
\ref{met_mass_age}c.  The separation between these two populations is inconsistent with being an
effect of the age--metallicity degeneracy in our fits, which primarily moves galaxies along the
trends in Fig. \ref{met_mass_age}.  If we consider fits to the emission- and non-emission-line
galaxies separately, both populations show a similar dependence on stellar mass; the offset between
these two populations is dominated by a varying age--metallicity relation.  Low-mass galaxies in our
sample exhibit an increase in scatter about the fitted plane relative to their high-mass
counterparts, in agreement with the observations of Smith \etal (2008b).

\begin{figure}
\centering
\includegraphics[scale=0.60,angle=-90]{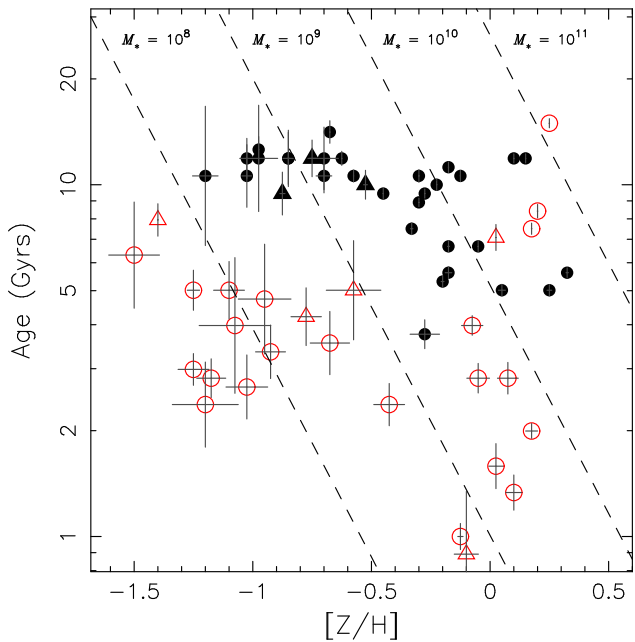}
\caption{Age--metallicity relation for NGC\,5044 group galaxies.  Symbols are the same as in Fig.
\ref{mass_sp}.  Lines of constant stellar mass are shown for $\log M_{\mathrm{*}} =
8,9,10,11\,M_{\mathrm{\odot}}$. Error bars are the 1-sigma errors determined from Monte Carlo
simulations (see $\S$\ref{fitting}).} 
\label{amr}
\end{figure}

The projections shown in Fig. \ref{met_mass_age} are helpful to examine the distribution of galaxy
parameters relative to one another, but are somewhat difficult to interpret in physical
terms.  In Fig. \ref{amr} we show a more standard projection of the plane in terms of age and
metallicity, where {\it dashed} lines represent the age--metallicity relation at fixed stellar mass
as derived from the best-fit plane discussed above.  

If we naively interpret the distribution of galaxy ages and metallicities shown in Fig. \ref{amr},
then there is is little evidence for a relationship between central galaxy age and metallicity;
galaxies formed at a similar time in the early universe exhibit a spread in metallicity of two or
more dex.  However, our fits in the age--mass--metallicity plane suggest that this is a flawed
conclusion.  While it is true to say that there is no clear age--metallicity trend when considering
our total sample of galaxies, subsamples of data in narrow mass bins (one or two orders of
magnitude) suggest that galaxies with older central ages are more metal poor than centrally young
galaxies, supporting evolution of the mass--metallicity relation over time (e.g. Kobulnicky \etal
2003; Brooks \etal 2007; Lamareille \etal 2007; Maiolino \etal 2008).

\subsection{Predicted vs. observed colours}
\label{pred_col}

\begin{figure*}
\centering
\includegraphics[scale=0.63,angle=-90]{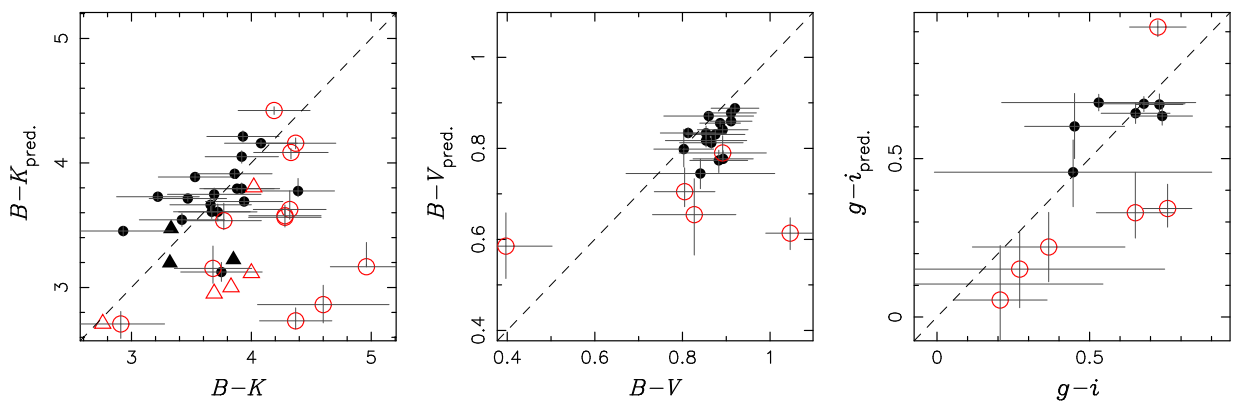}
\caption{BC03 predicted colours vs. observed global $B-K$, $B-V$ and $g-i$.  Filled and open symbols
represent passive and emission-line galaxies respectively.  Errors on predicted colours are
calculated using the errors estimates on our central ages and metallicities.  {\it Dashed} lines
represent equality between the predicted and observed colours.} 
\label{pred_obs}
\end{figure*}

While up to this point we have used the BC03 models to derive M/L, and hence mass, for our galaxies,
it is important to establish the extent to which these SSP models accurately describe the properties
of our sample galaxies.  To undertake this comparison we have supplemented the $B$-- and $K$--band
photometry used in Paper I with $B-V$ and $g-i$ colours for a subsample of NGC\,5044 group dwarf
galaxies presented by Cellone (1999) and Cellone \& Buzzoni (2001; 2005; S. Cellone, private
communication).  Predicted colours are calculated using the BC03 models and measured ages and
metallicities.  Comparisons of these predictions to our observed colours are shown in Fig.
\ref{pred_obs}.

Galaxies lacking in emission show relatively good agreement with the predicted BC03 colour for their
age and metallicity.  This serves as an excellent confirmation in the reliability of M/L estimates
for these galaxies which, to some extent, can be seen from the tightness of the mass-metallicity
relation in Fig. \ref{mass_sp}.  The scatter in colours of emission line galaxies is considerable,
and the causes of this are likely to be two-fold: firstly, we are sampling fundamentally different
regions with our spectroscopy and photometry.  Whereas this has a negligible effect for passive
galaxies, the centrally concentrated star-formation in our emission galaxies means that the colours
we predict should be, and indeed are, generally bluer than the observed ``global'' photometry.  In
addition, stellar population parameters derived for our emission-line galaxies are the most
uncertain, which leads to a greater uncertainty in both our measurements of ages and metallicities
and hence a corresponding uncertainty in their predicted colours.  Overall, however, the above
comparisons suggest that our galaxy data are reasonably well described by the BC03 models.

\section{Global properties and galaxy distribution}
\label{group_params}

We have so far examined general trends in the galaxy population of the NGC\,5044 group in their own
right.  We now turn to a more general discussion of the distribution of galaxy properties within the
group.

\subsection{Stellar population trends with radius}
\label{pop_radius}

By combining semi-analytic models for star formation and galaxy evolution with recent, large-scale
$N$-body simulations such as the Millennium Simulation (Springel \etal 2005), recent theoretical
work has made great strides in predicting the distribution of galaxy properties in massive
structures like groups and clusters.  As an example, de Lucia \etal (2006) found that the
luminosity-weighted age, metallicity and stellar mass fall with increasing distance from the cluster
centre in their models, out to the virial radius.  Given the hierarchical formation scenario
favoured in current $\Lambda$CDM models these results are not surprising; those galaxies in the
highest density regions form first, and subsequent generations of galaxies accreted by the cluster
are distributed with radius according to the redshift at which they become cluster members (e.g. Gao
\etal 2004).  It should be noted, however, that there is considerable scope for the interpretation
of projected cluster-centric distance as an indicator of accretion time to be confused by three-body
interactions which can eject {\it bona fide} members to beyond the virial radius (see e.g. Ludlow
\etal 2008).

In light of these predictions, we are motivated to examine the radial distribution of stellar
populations in the NGC\,5044 group.  Galaxies residing in groups and clusters for more than a
dynamical time should undergo some degree of mass segregation due to the increasing efficiency of
dynamical friction with total galaxy mass.  In apparently relaxed groups and clusters, such as
NGC\,5044, we expect any such segregation to be readily apparent.  In Paper I we examined the
distribution of dwarf and giant galaxies in the NGC\,5044 group using $B$-band magnitudes as a
simple discriminator between low- and high-mass systems but found no significant evidence for
differing radial distributions of these two sub-populations.  Here, we are able to refine this
analysis using the stellar masses derived from the BC03 models.  

\begin{figure*}
\centering
\includegraphics[scale=0.70,angle=-90]{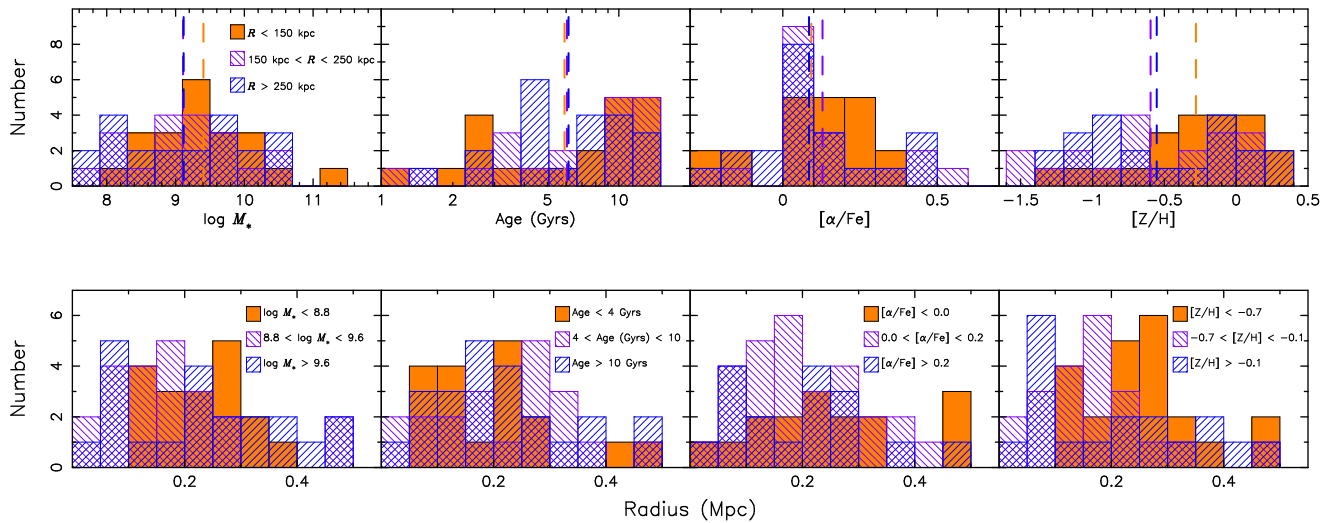}
\caption{The radial dependence of various properties of galaxies in the NGC\,5044 group.  Top panels
show the distribution of stellar mass, age, $\alpha$-element abundance and metallicity in three
different radial bins of $R < 150$\,kpc, $150$\,kpc $\leq R \leq 250$\,kpc and $R > 250$\,kpc (solid
orange, green and blue hatched regions respectively).  Vertical {\it dashed} lines show the mean of
each bin.  Lower panels show the radial distribution of galaxy properties in three separate bins of
stellar mass, age, $\alpha$-element abundance and metallicity, where bins have been selected to
contain roughly equal numbers of galaxies.}
\label{prop_hist}
\end{figure*}

In the upper panels of Fig. \ref{prop_hist} we show the distribution of stellar mass, age,
$\alpha$-element abundance and metallicity in fixed radial bins.  In the lower panels we show the
radial distribution of galaxies in bins of stellar mass, age, $\alpha$-element abundance and
metallicity.  This figure includes only those galaxies within 500\,kpc of the group centre (roughly
two-thirds the group's virial radius) as this is approximately the region observed uniformly with
our new AAOmega observations, and thus less prone to spurious, selection-induced trends.  

Focusing first on the distribution of properties in fixed radial bins, we find no significant
differences in the mean values for age or $\alpha$-element abundance (shown by vertical {\it dashed}
lines in the top panels of Fig. \ref{prop_hist}).  Data show some evidence for a deviation in mean
galaxy stellar mass and metallicity, and a Kolmogorv-Smirnov (KS) test confirms that metallicities
in the inner 150\,kpc of the group are inconsistent with being drawn from the same distribution as
either of the outer two radial bins at the 2.5 sigma level.  Despite the apparent offset in mean
stellar mass, a KS test does not find galaxies in the inner group region to have a significantly
different mass distribution relative to other radial bins.  

Turning to the radial distribution of galaxy properties (Fig. \ref{prop_hist}, lower panels), a KS
test finds that both low-mass and low-metallicity galaxies ($\log M_* < 8.8$ and [Z/H]$ <
-0.7$\,dex) exhibit radial distributions that are different to their higher mass and metallicity
counterparts.  Qualitatively, these differences are evident in Fig. \ref{prop_hist} by the clear
lack of galaxies from the lowest mass and metallicity bins in the inner 100\,kpc of the group
centre.  The fact that we see this segregation in both mass and metallicity is expected from the
tight correlation between these two quantities in our data; the converse holds true for the lack of
peculiar distributions in age and $\alpha$-element distributions given their lack of a strong
correlation with mass.

\begin{figure}
\centering
\includegraphics[scale=0.65,angle=-90]{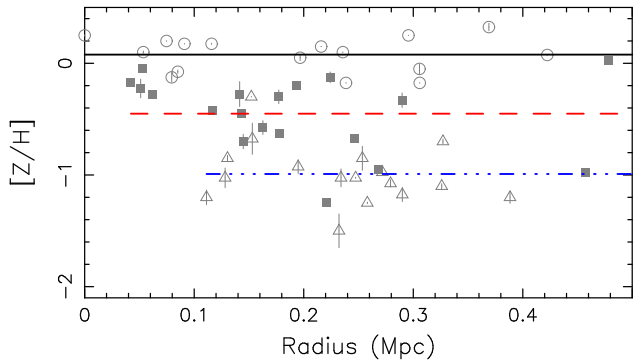}
\caption{Galaxy metallicity against projected group-centric radius.  Galaxies are divided into three
mass bins with approximately equal numbers such that circles, squares and triangles represent
galaxies with $\log M_{*} > 9.6$, $8.8 \leq \log M_{*} < 9.6$ and $\log M_{*} \leq 8.8$.  {\it
Solid}, {\it dashed} and {\it dot-dashed} lines represent the mean metallicity for each mass bin.}
\label{rad_mass}
\end{figure}

In order to investigate the relationship between mass, metallicity and radius further, in Fig.
\ref{rad_mass} we show galaxy metallicities plotted against their group-centric radii, where
galaxies have been separated into three separate mass bins.  This figure serves as a visual
confirmation of the mean offset of galaxy metallicity observed within $\sim$150\,kpc in Fig.
\ref{prop_hist}, and also shows that this trend is due primarily to a lack of low-mass, and hence
low-metallicity, galaxies from the inner 100\,kpc of the group.  Finally, Fig. \ref{rad_mass}
suggests that there are no strong radial trends in metallicity at fixed stellar mass. 

Previous work by Mathews \etal (2004; hereafter M+04) using the NGC\,5044 group catalogue of
Ferguson \& Sandage (1990; hereafter FS90) has noted a similar lack of low-luminosity galaxies in
the inner group region.  Assuming dwarf galaxies act as tracer particles in the group potential,
M+04 compared the cumulative number distribution of dwarf galaxies to the predicted surface mass
distribution of a Navarro-Frenk-White (NFW) dark matter profile (Navarro, Frenk \& White 1995, 1996,
1997), finding that the NGC\,5044 group falls well below the NFW prediction inside $\sim$300\,kpc.
M+04, and later Faltenbacher \& Mathews (2005; hereafter FM05), offer two explanations for this
apparent central deficit of dwarf galaxies.  M+04 argue that a lack of dwarf galaxies could be tied
to a low survival rate of low-mass galaxies entering the NGC\,5044 group halo at high redshift ($2-3
\leq z \leq 6$), or due to suppression of star formation in low-mass galaxies as a result of AGN or
outflow activity from the central galaxy.  The follow-up work by FM05 showed, using a simple
dynamical model, that tidal disruption and dynamical friction could also be responsible for the
apparent lack of dwarfs at small radii.  In their model, low-mass galaxies approaching the central
group potential could be strongly disrupted or destroyed.  

Our sample has allowed us to refine the NGC\,5044 group membership used by M+04, and we find that
limiting the FS90 group samples to only the spectroscopically defined members in Paper I does not
significantly alter the conclusions of M+04\footnote{Where the NFW profile is scaled to match the
cumulative galaxy distribution at $\sim$350\,kpc as in M+04}; there still appears to be a deficit of
dwarf galaxies at low projected radii ($\la$300\,kpc).  M+04 suggest that, as a result of low dwarf
galaxy survival rates at high redshifts, there should be a relative lack of old, low-mass galaxies
in the group.  The data presented in Fig. \ref{prop_hist} do not suggest any such trend in galaxy
ages, with low-mass galaxies appearing similar in mean age to their high-mass counterparts (see also
Figs.  \ref{mass_sp} and \ref{amr}).  In addition, the M+04 scenario would suggest that, relative to
massive galaxies, low-mass dwarf galaxies should be kinematically ``younger'' in their velocity
distribution in the group; in Paper I we find no such evidence for kinematic segregation between
high- and low-luminosity galaxies.  A repeat analysis of sub-population kinematics using the stellar
masses derived in this work gives a similar result, further suggesting that the scenario posed by
M+04 does not describe these data in full.

Perhaps the most plausible explanation is that put forward by FM05, where the apparent central
deficit of dwarf galaxies is a result of both dynamical friction and tidal disruption.  In their
work, FM05 find that the evolution of galaxies with stellar to total mass ratios of $\sim$20 best
describe the population of disrupted galaxies in the NGC\,5044 group.  Using the relationship
between stellar mass and total baryonic mass determined from fits to star-forming galaxies in the
SDSS (Baldry, Glazebrook \& Driver 2008), we can use the cosmic baryon fraction, $f_b = 0.171$ as
determined from the {\it WMAP} 5yr results (Komatsu \etal 2008) to estimate the total halo mass
for galaxies of a given stellar mass.  For the stellar-to-total mass ratio predicted by FM05 of 20,
the above above relation gives a corresponding stellar mass of $\log M_* \approx 8.6$.  This is in
agreement with the lower limit of stellar mass that we observe in the central 100\,kpc of the group
centre, providing support for the tidal disruption scenario of FM05.

Confirmation of this would need to come from additional deep imaging to observe the low
surface-brightness intragroup light which, given the possibly large number of disrupted or destroyed
dwarf galaxies, could be as much as $\sim$35 percent of the group's total luminosity (FM05).

\subsection{Galaxies and the hot intragroup medium}
\label{gas}

The metal content of the intracluster/intragroup medium (ICM/IGM) is directly related to the star
formation histories of the galaxies via outflows and supernovae driven winds, and therefore provides
a useful diagnostic of both group and galaxy evolution.  Despite the presence of ongoing evolution
in the stellar populations of group and cluster galaxies, observations of high redshift X-ray
samples suggest that the bulk of ICM metals were already in place at $z\sim1$ (e.g.  Tozzi \etal
2003; Balestra \etal 2007) and therefore necessitate that {\it most} star formation activity and
subsequent ICM enrichment takes place at early times.   

\begin{figure}
\centering
\includegraphics[scale=0.40,angle=-90]{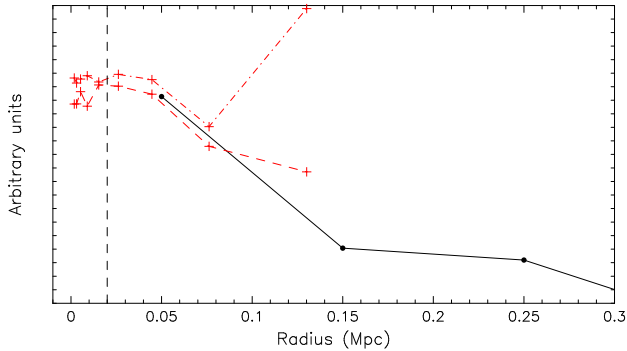}
\caption{Comparison of stellar mass density ({\it solid} black line) and X-ray derived iron and
silicon abundances (red {\it dashed} and {\it dot-dashed} lines respectively) as a function of
radius.  Data have had a fixed offset applied to be shown on the same figure. X-ray data have been
binned for clarity.  The vertical {\it dashed} line represents the radial extent of NGC\,5044's
light profile of $\sim$20\,kpc.} 
\label{mass_profile}
\end{figure}

Rasmussen \& Ponman (2007; hereafter RP07) have used {\it Chandra} archival data to analyse the
X-ray abundance properties of 15 nearby groups, including NGC\,5044.  From their data RP07 extract
radial profiles of iron and silicon abundance in the hot IGM out to $\sim200$\,kpc in NGC\,5044,
facilitating a comparison between our galaxy stellar population measurements and the chemical
properties of the intragroup gas.  In Fig. \ref{mass_profile} we plot the X-ray derived radial iron
and silicon abundances along with the stellar mass-density profiles.  

In their abundance analysis of hot gas in the NGC\,5044 group, RP07 show that the central silicon to
iron ratio is roughly solar (i.e. [Si/Fe]$\sim0.0$), suggesting SNIa are playing an important role
in enriching gas in the central group regions.  While we lack significant overlap between the X-ray
profiles and our binned group profile, the slopes in both iron and silicon abundance in the
intragroup gas are consistent with that of the stellar mass density between 50 and $\sim$100\,kpc.
In clusters, the central ``excess'' of iron in the central regions relative the rest of the ICM is
normally attributed to enriched outflows from the central galaxy (e.g. B\"ohringer \etal 2004).  In
the case of NGC\,5044 this remains a plausible explanation for the high central iron abundance,
particularly given recent evidence for large outflows associated with NGC\,5044's central AGN (Temi,
Brighenti \& Mathews 2007; Gastaldello \etal 2007).  We note, however, that our data are also
consistent with satellite galaxies having contributed significantly towards building up the central
iron peak in the group, particularly given the extended nature of the central excess, $\sim$50\,kpc,
relative to the optical extent of NGC\,5044 ($\sim$20\,kpc; FS90; Paturel \etal 2000; shown with
black {\it dashed} line in Fig. \ref{mass_profile}) and the the good correspondence between the
radial profile of iron abundance and stellar mass-density out to beyond $\sim$100\,kpc.

In the outer regions, $R>100$\,kpc, there is evidence for a rising silicon abundance, in contrast to
the iron abundance which continues to fall.  While the significance of this upturn is only marginal
due to its dependence on the outer-most measurement in the X-ray data, we have no reason to believe
this is a spurious measurement.  The observed rise in [Si/Fe] is in agreement with X-ray
observations for numerous groups and clusters (e.g. Finoguenov \etal 2000; RP07) and suggests an
increased contribution of SNII relative to SNIa at large projected radii.

\begin{figure}
\centering
\includegraphics[scale=0.62,angle=-90]{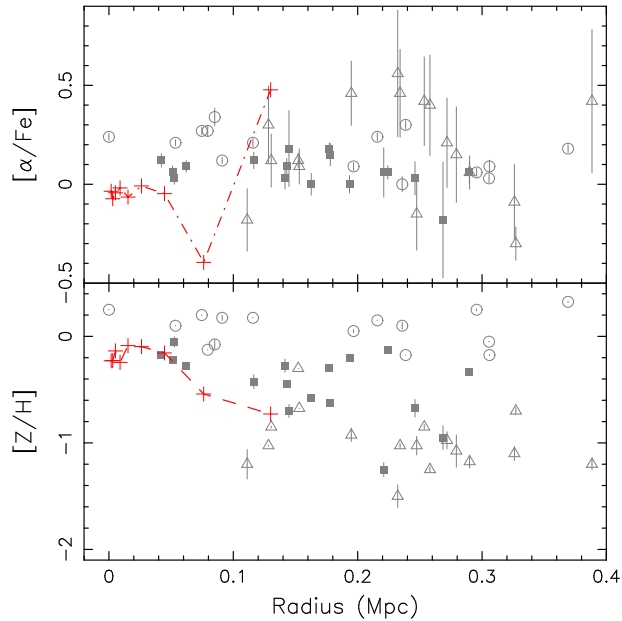}
\caption{Comparison of X-ray derived silicon and iron abundance profiles ({\it dashed} lines in each
panel) with the distribution of galaxy $\alpha$-element abundances and metallicities.  Galaxy
symbols are the same as in Fig. \ref{rad_mass}.  X-ray data have been binned for clarity.} 
\label{xray_profile}
\end{figure}

To explore the relation between galaxies and the intragroup medium further, in Fig.
\ref{xray_profile} we show the same binned X-ray abundance profiles as in Fig. \ref{mass_profile},
only this time overlayed on the radial galaxy trends for metallicity and [$\alpha$/Fe] shown in Fig.
\ref{rad_mass}.  This figure clearly shows that both the IGM iron abundance and the [Si/Fe] ratio
are well below the average values measured in galaxies at similar radii, despite both iron and
silicon declining with radius at a similar rate to the stellar mass-density. On the other hand,
low-mass galaxies generally exhibit stellar metallicities which fall below the measured IGM iron
abundance at a given radius.  Given the shallow potential wells of low-mass galaxies,
metal-enhanced starburst or supernovae driven winds provide a plausible explanation for the observed
offset (see also Section \ref{sp_params}).  Low-mass galaxies not withstanding, the observed low
chemical abundance of the IGM is consistent with the findings of Buote, Brighenti \& Mathews (2004),
who also note the low abundance of iron in the IGM relative to the group luminosity (compared to
clusters), as well as other work suggesting a high fraction of ``primordial'' gas in the IGM (e.g.
Gibson \& Matteucci 1997; Moretti, Portinari \& Chiosi 2003).   

In terms of silicon abundance, we find a surprising correlation between the rise in [Si/Fe] at
$\sim$100\,kpc and the presence of low-mass, dwarf galaxies at similar radii.  If there is a
connection between these two observations, it would imply a very high contribution of SNII relative
to SNIa in dwarf galaxies, as well as a significant contribution from these low-mass systems to the
total gas mass of the group.  Previous studies, however, have shown that, while dwarf galaxies
certainly contribute to the reservoir of intragroup gas, it is most likely not significant,
constituting {\it at most} 15\% of the total gas mass in clusters (Gibson \& Matteucci 1997).  This
suggests that the connection between these observations is likely only coincidental, particularly
given the relatively low IGM iron abundance outside the group core, 10\% solar, and the likely
sub-solar outflows from dwarf galaxies.

\section{Emission line properties}
\label{emm_line}

Using the sub-sample of our NGC\,5044 group with measurable emission lines, we now discuss the
emission-line characteristics of our group galaxies, including star-formation rates and nebular
metallicities.

\begin{figure}
\centering
\includegraphics[scale=0.48,angle=-90]{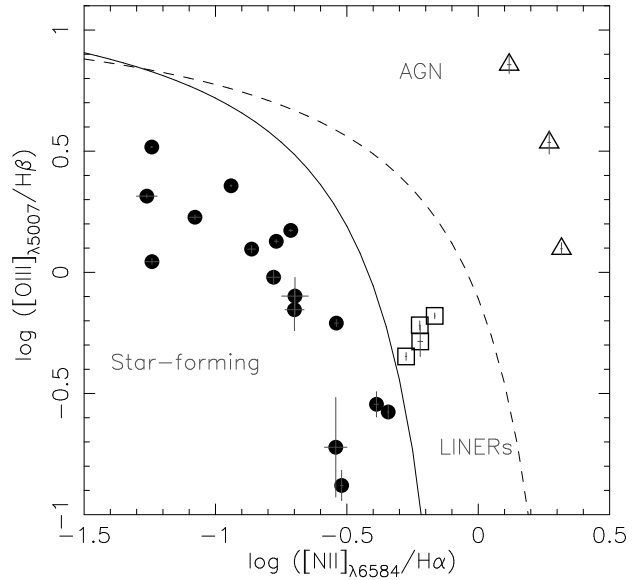}
\caption{Line diagnostic for emission galaxies in our NGC\,5044 group sample.  The {\it solid} and
{\it dashed} lines represent the semi-empirical pure star formation line from Kauffmann \etal (2003)
and the theoretical maximum starburst line from Kewley \etal (2001) respectively.  Symbols represent
galaxies which are likely purely star-forming, AGN or composite AGN and star-forming (circles, open
triangles and open squares respectively).} 
\label{bpt}
\end{figure}

First, in order to separate galaxies undergoing star formation activity from those with other strong
ionising sources, i.e. AGN, we use a standard emission line diagnostic comparing the
[OIII]$_{\lambda5007}$/H$\beta$ and [NII]$_{\lambda6584}$/H$\alpha$ flux ratios (Baldwin, Phillips
\& Terlevich 1981), shown in Fig. \ref{bpt}.  We use two predictions to characterise the emission
line flux ratios of our galaxies.  The first is the theoretical maximum starburst limit calculated
by Kewley \etal (2001; {\it dashed} line in Fig.  \ref{bpt}), while the second is the semi-empirical
limit of pure star formation defined by Kauffmann \etal (2003) using a large sample of SDSS emission
line galaxies ({\it solid} line in Fig. \ref{bpt}).  These lines separate our sample into galaxies
which are purely star-forming, strongly AGN or ionised by a composite source (some contribution from
both AGN and star-formation), shown with circles, triangles and squares in Fig. \ref{bpt}. 

NGC\,5044 is known to host AGN activity (Rickes, Pastoriza \& Bonatto 2004, Brough \etal 2007) and
is identified as such in Fig. \ref{bpt}, but we also identify two other group galaxies as hosts
of potential AGN or other strong ionising sources: NGC\,5037 (FS 068) and FS 082.  The
classification of these galaxies as AGN sources is supported by the available {\it Chandra} data, in
which all three galaxies are detected.

\subsection{Star formation rate}
\label{sfr}

In large samples, comparisons of galaxy properties with local environment show strong evidence for a
decreasing fraction of star-forming galaxies in regions of higher projected galaxy surface density
(Lewis \etal 2002; G\'omez \etal 2003; Poggianti \etal 2006).  Even at relatively low gas densities,
such as those found in groups, recent work has shown that ram-pressure stripping can influence
star-formation via depletion of a galaxy's hot-gas reservoir (Sivakoff \etal 2004; Machacek \etal
2005; Rasmussen \etal 2006; Kawata \& Mulchaey 2008).  However in higher-density environments, such
as the Coma cluster, Poggianti \etal (2004) have found evidence for young post-starburst galaxies
preferentially located near the edges of X-ray substructures.  These observations suggest that
galaxy--ICM interactions may play a key role in the truncation of star formation, but may equally
contribute to initiating bursts of star formation in galaxies as they encounter dense environments. 

In our own data, varying fibre throughput and sensitivity make proper flux calibration of our data
difficult and therefore limit our ability to derive absolute measurements of the star formation
rate, SFR, using H$\alpha$ line fluxes.  Instead, we use the relation derived by Guzm\'an \etal
(1997) to estimate star formation rates using $B$-band luminosity and [OII] equivalent widths, where  

\begin{equation}
\label{o2_sfr}
\mathrm{SFR(}M_\mathrm{\odot}\mathrm{yr^{-1})} \approx 2.47\times10^{-12}\,L_\mathrm{B}(L_\mathrm{\odot})\,\mathrm{EW_{[OII]}}.
\end{equation}

\noindent  We use the $B$-band luminosities for galaxies as published in Paper I, which are taken
from FS90 and Paturel \etal (2005).  

The use of [OII] instead of H$\alpha$ introduces an additional error into the estimation of the star
formation rate of order 0.2 to 0.3\,dex (Kewley \etal 2004).  The relation fit by Guzm\'an \etal
(1997) to convert between [OII] flux and [OII] equivalent width carries with it increased
uncertainty of $\sim$0.1\,dex in the derived SFR.  Given these two relatively significant error
contributions, in addition to the relatively large errors in the FS90 and Paturel \etal photometry
(0.5 and 0.3\,dex respectively, see Paper I), we consider our star formation rates to only be
proportional to the star formation activity in any given galaxy, but by no means a clear indicator
of the absolute star formation rate.  

\begin{figure}
\centering
\includegraphics[scale=0.35,angle=-90]{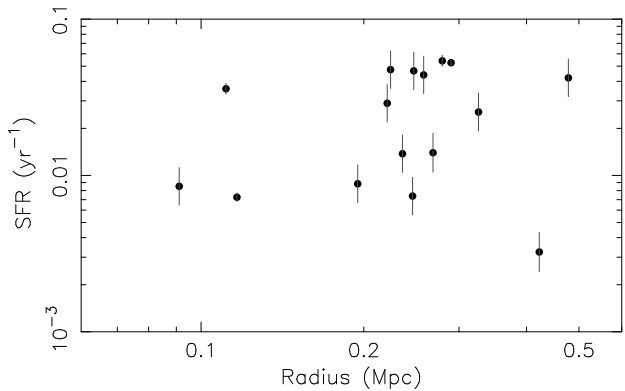}
\caption{Specific star formation rate as it varies with projected distance from the group centre.
Errors are indicative of the uncertainty in our EW$_{\mathrm{[OII]}}$ and $B$-band luminosity
measurements, but do not include the systematic contribution from using [OII] over H$\alpha$ and
equivalent widths as opposed to line fluxes (see $\S$\ref{sfr}).}
\label{sfr_rad}
\end{figure}

In Fig. \ref{sfr_rad} we show the specific star formation rate for galaxies as estimated using Eqn.
\ref{o2_sfr} in relation to their projected radius from the group centre.  The star formation rates
shown in Fig. \ref{sfr_rad} have been normalised by the galaxy stellar mass and are shown in units
of yr$^{-1}$ per 10$^9$\,$M_{\mathrm{\odot}}$.  We see no clear correlation of SFR with radius, and
results are similar if we consider projected galaxy density.  Motivated by the possible dependence
of SFR on galaxy--ICM/IGM interaction, we have also examined the star formation rate relative to a
simple expression proportional to the IGM ram pressure, $P_{ram} =
\rho(r)|\bar{\nu}_{\mathrm{group}}-\nu|^2$, where $\rho(r)$ is the IGM density and
$|\bar{\nu}_{\mathrm{group}}-\nu|$ is the galaxy velocity $\nu$ relative to systemic group velocity
$\bar{\nu}_{\mathrm{group}}$.  Assuming that $\rho(r) \propto r^{-2}$, we then arrive at a
comparison of SFR against $|\bar{\nu}_{\mathrm{group}}-\nu|^2/r^2$, which shows a similar lack of
correlation as the radial and galaxy density comparisons described above.  We note that projection
effects are likely having a strong effect on any of our radial SFR comparisons as we cannot properly
relate our galaxy positions and recession velocities to their true position and relative motion
withing the group.  Nevertheless, while these data do not allow us rule out the role of interactions
with intragroup gas as influencing the observed star formation in these galaxies, they seem to hint
that these effects are relatively limited in this case.  
 
An alternative possibility for triggering or enhancing star formation is through galaxy--galaxy
interactions.  Hydrodynamic simulations modelling gas have shown that tidal interactions can enhance
or induce central star formation by driving gas from the disk to the central regions (e.g. Barnes \&
Hernquist 1992; Di Matteo \etal 2007), which supports observations of enhanced star formation in
close pairs of galaxies (e.g. Lin \etal 2007; Ellison \etal 2008).  An examination of DSS images for
the 22 galaxies with [OII] emission shows that only 2 galaxies, FS90 134 and FS90 137, exhibit clear
evidence for ongoing interaction.  Several other galaxies show some evidence of disturbed morphology
(i.e.  boxy bulges and disturbed disks), however it appears that the majority of recent star
formation activity we see in the NGC\,5044 group is not driven by currently ongoing galaxy--galaxy
interactions.

\subsection{Emission-line metallicity}

Nebular metallicity and underlying, stellar metallicity are intimately related via star formation.
The evolution of stellar metallicity (as measured by absorption features) is strictly dependent on a
galaxy's star formation history, while nebular metallicities continue to vary as previous
generations of stars evolve and enrich the interstellar medium.  We are therefore able to use the
relative offset of emission- and absorption-line metallicities to extract information related to the
last significant star formation episode.  Here we use a reparameterisation of the Kewley \& Dopita
(2002; hereafter KD02) $R_{23}$ method presented by Kobulnicky \& Kewley (2004; hereafter KK04) to
derive gas-phase metallicities for a subsample of our emission-line galaxies.  We exclude from this
analysis galaxies exceeding the maximum starburst limit shown in Fig. \ref{bpt} as likely hosting
AGN (NGC\,5044, NGC\,5037 and FS\,082). 

The $R_{23}$ metallicity calibrator is typically defined using emission line fluxes, however this
requires spectra to be properly flux calibrated as line ratios spanning a significant wavelength
range will otherwise be incorrect.  An alternate approach is to use measurements of equivalent width
in place of line fluxes to compute EW-$R_{23}$, which avoids the issues of flux calibration and
extinction correction, but introduces the problem of potential variation in the underlying stellar
continua (particularly between old and young spectra).  In a statistical sense, the variation
between $R_{23}$ and EW-$R_{23}$ results in an an additional error on the final metallicity
determination of $\sim$0.11\,dex (Kobulnicky \& Phillips 2003), however in our relatively small
sample where we {\it know} the underlying stellar populations vary significantly, the uncertainty
between $R_{23}$ and EW-$R_{23}$ can have a potentially significant effect on our results.
Considering both the wide range of spectra types and uncertain flux calibration of our data, we
estimate nebular metallicities using both $R_{23}$ and EW-$R_{23}$.  Since the effects of varying
stellar continua and flux calibration affect emission-line metallicity measurements differently, by
using both of these measurements we can assess the reliability of our conclusions.

Comparison of EW-$R_{23}$ and $R_{23}$ shows that the scatter between the two measures is
$\sim$0.08\,dex, or $\sim$0.11\,dex in terms of the final determination of $12+\log(\mathrm{O/H})$
(Kobulnicky \& Phillips 2003), which is generally less than the intrinsic error in the calibration
of $R_{23}$ to nebular oxygen abundance.  In our own data, the scatter between EW-$R_{23}$ and
$R_{23}$ is $\sim0.16$\,dex, but this scatter is primarily driven by two galaxies with
significant offsets between EW-$R_{23}$ and $R_{23}$.  Excluding these galaxies the scatter on our
data falls to $\sim$0.07\,dex.  We therefore exclude these deviant galaxies from the remainder of
our emission line analyses and consider measurements on the remaining galaxies to be reliable.

\begin{figure}
\centering
\includegraphics[scale=0.53,angle=-90]{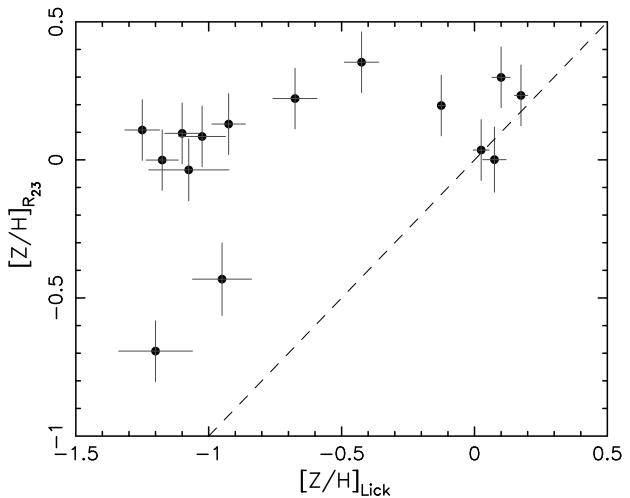}
\caption{Comparison of metallicity derived from the nebular oxygen abundance,
[Z/H]$_{\mathrm{R_{23}}}$ and stellar metallicity, [Z/H]$_{\mathrm{Lick}}$, derived using Lick
indices.  The {\it dashed} line represents equality between the nebular and stellar metallicities.}
\label{met_comp}
\end{figure}

Comparison of stellar and nebular metallicities is carried out by converting nebular oxygen
abundances to the implied metal mass fraction using the relation $Z \simeq
29\times10^{[12+(O/H)]-12}$, assuming the standard solar abundance distribution and solar oxygen
abundance of $12 + \mathrm{(O/H)} = 8.72$ (Allende Prieto, Lambert \& Asplund 2001; Kobulnicky \&
Kewley 2004).  In Fig.  \ref{met_comp} we show the comparison between the converted nebular
metallicities and stellar metallicity.  We find that, in general, derived nebular metallicities are
super-solar, and in some cases nearly 1\,dex higher than their stellar counterparts.  The relatively
shallow relation between stellar and nebular metallicities is somewhat puzzling, as it implies a
significantly weaker mass-metallicity relation among star-forming galaxies than is generally
observed (e.g. Tremonti \etal 2004).

\begin{figure*}
\centering
\includegraphics[scale=0.60,angle=-90]{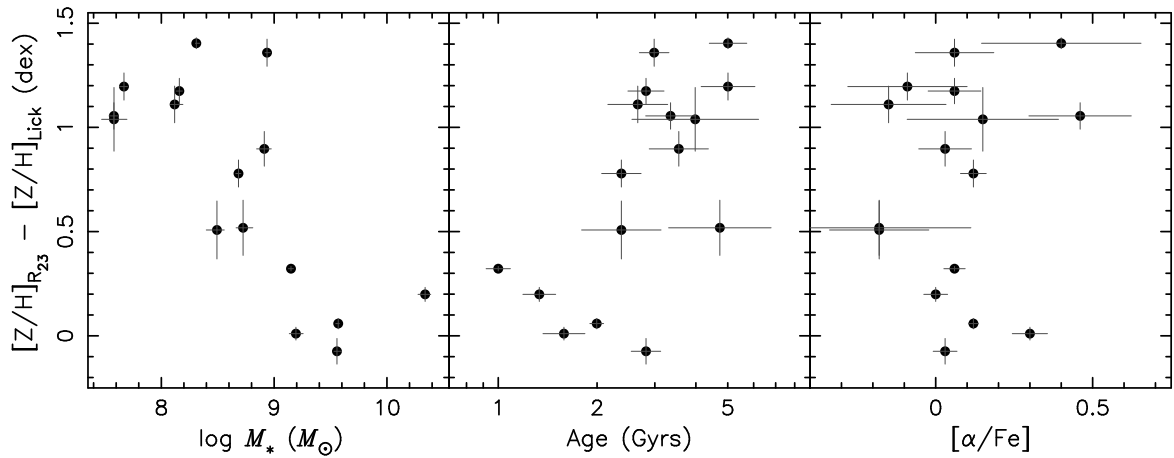}
\caption{Offset between nebular and stellar metallicity plotted against stellar mass, age and
$\alpha$-element abundance.} 
\label{gas_offset}
\end{figure*}

Given this somewhat curious result we are motivated to examine the relative difference between
nebular and stellar metallicity as a function of other galaxy properties.  In Fig. \ref{gas_offset},
the relative offset between nebular and stellar metallicity is plotted against galaxy stellar mass,
age and $\alpha$-element abundance.  The relationship between metallicity offset and stellar mass in
Fig. \ref{gas_offset} is driven primarily by the mass--metallicity relation and shows that the most
massive galaxies have the lowest measured nebular--stellar metallicity offsets as a result of the
relatively uniform, high nebular metallicities measured.  

In terms of age, those galaxies with the lowest metallicity offsets also have the youngest
luminosity-weighted ages, which is consistent with the observed stellar population forming recently
from gas present in these galaxies.  Interestingly, while the trend of metallicity offset with age
is consistent with our understanding of galaxy formation, the magnitude of this offset is large,
nearly 1.5\,dex in the case of the lowest-mass galaxies.  This offset is difficult to explain
through pure passive evolution of the stellar populations in these galaxies, i.e. through enrichment
from SNII and SNIa, which in extreme cases can account for $\sim$1\,dex of metallicity increase
(see, e.g. Sansom \& Proctor 1998).  An alternate explanation is that {\it ongoing} star formation
has helped to enhance the observed gaseous metallicities through the evolution of metal-enriched
stars.  In this case the integrated stellar populations should also appear both younger and more
metal rich in those galaxies hosting the largest nebular--stellar metallicity offsets, in conflict
with the trends in Fig. \ref{gas_offset}. Finally, these data could be explained in a scenario where
the gas reservoirs of low-mass galaxies are being preferentially depleted over high-mass galaxies,
which would increase the effect of any subsequent enrichment due to supernovae.  

In order to explain these observations in terms of gas removal, we require that the removal process
must be more efficient in low-mass galaxies, ruling out AGN activity as a possibility due to its
strong scaling with mass.  Starburst outflows, primarily driven by SNII, satisfy the above stated
mass dependence but would also likely affect the observed [$\alpha$/Fe] ratio of stars formed from
any remaining gas, in conflict with Fig. \ref{gas_offset}c.  Galaxy--IGM interactions such as
ram-pressure stripping (Gunn \& Gott 1982) and viscous stripping (Nulsen 1982) satisfy both the
inverse mass dependence and non-preferential gas removal (i.e. gas-phase metals from SNII and SNIa
must be removed more-or-less equally) stipulated by these observations.  Strictly speaking, we
cannot rule out any processes from contributing to the observed galaxy processes, however our data
are best explained by galaxy--IGM dominating gas removal in the NGC\,5044 group.

\section{Discussion and conclusions}
\label{conclusions}

In this second of two papers, we have undertaken a spectroscopic investigation of galaxies in the
group environment, deriving stellar masses, ages, metallicities and $\alpha$-element abundance
ratios for 67 of the 111 spectroscopically confirmed NGC\,5044 group members described by Mendel
\etal (2008).  These measurements have allowed us to examine the star formation histories of these
group members, both as a population in their own right and in the context of the group environment
and their location in it.

The mass--metallicity relation plays a fundamental role in describing galaxy populations, and we see
strong evidence of this in the overall tightness of the mass--metallicity relation in our group data
over multiple orders of magnitude in stellar mass.  In the context of environment, by comparing the
mass--metallicity relation of NGC\,5044 group galaxies with galaxies in the local group and the
Shapley supercluster, we show that galaxies appear to form a continuous mass--metallicity relation
across several orders of magnitude in system mass and spanning upwards of six orders of magnitude in
galaxy stellar mass.  These data argue for a relative independence of the mass-metallicity relation
from environment (e.g. Maiolino 2008; van den Bosch \etal 2008).

In terms of galaxy ages and $\alpha$-element abundances, we find no evidence for strong trends with
galaxy mass.  Our $\alpha$-element abundance data do not rule out the interpretation of star
formation efficiency as the primary driver of the mass--metallicity relation.  Particularly at
higher stellar masses ($\log M_* \ga9\,M_{\odot}$) and metallicities, where wide separation of model
grids allow for finer abundance measurements, our data show hints of a trend of increasing
$\alpha$-element ratio with mass, consistent with literature results (e.g. Terlevich \& Forbes 2002;
Thomas \etal 2005).  We have shown that the surface described simultaneously by age, mass and
metallicity is vital to analyses of galaxy data spanning a broad range in mass.  Most notably, mass
dependence is vital to interpretation of the galaxy age--metallicity relation.

In terms of the galaxy radial distribution, we find no apparent radial dependence in either age or
$\alpha$-element abundances.  A KS test finds that low-mass and low-metallicity galaxies have radial
distributions inconsistent with their high-mass and high-metallicity counterparts, and an
examination of this apparent difference shows that we observe {\it no} galaxies with $\log M_*
\la8.9\,M_{\odot}$ within $\sim$100\,kpc projected radius from the group centre.  In the context of
previous analyses of the NGC\,5044 group, M+04 and FM05 have both noted the apparent lack of dwarf
galaxies in the group's central regions using the photometric catalogue of FS90.  Our spectroscopic
observations confirm that the dwarf galaxy distribution observed by M+04 and FM05 is not due to the
inclusion of significant foreground or background contamination in the FS90 catalogue (see Paper I).
Of the possible scenarios for explaining this deficit, the most plausible seems to be that dwarf
galaxies are being tidally disrupted or destroyed through interactions with the group or central
galaxy potential.  Our data are consistent with the simulations of FM05 in their prediction of
stellar-to-total mass ratios for disrupted galaxies of $\sim20$, corresponding to $\sim \log M_*
\approx 8.6$.  Deep wide-field imaging is required to confirm this conclusion, as the models of FM05
predict stars from the disrupted population of dwarf galaxies to contribute nearly 35 percent of the
total group luminosity, consistent with estimates of intracluster light for other groups at a
similar mass (e.g. Gonzalez, Zaritsky \& Zabludoff 2007).
 
With regard to similar types of analyses on other groups and clusters, Coma is the only cluster
analysed to a similar depth with stellar population measurements.  Smith \etal (2008a) have examined
radial gradients of Coma cluster dwarf galaxies, finding strong trends towards younger ages and
higher metallicities with increasing cluster-centric radii.  While these results are at odds with
our own, we believe this disagreement to be largely driven by the galaxy selection employed by Smith
\etal (2008a).  The trends observed by Smith \etal (2008a) are consistent with the interpretation
that the south-west ``infall'' region of Coma has already undergone a passage through the cluster
centre.  Observations of younger ages and higher metallicities in a ``quenched'' dwarf population
would then be consistent with the idea of induced star formation from galaxy encounters with the
dense/hot IGM (e.g.  Poggianti \etal 2004).  The fact that we do not see such a trend would seem to
suggest that we either fail to measure galaxies at sufficient radii to see effects of induced star
formation in our galaxies, or that such dramatic interactions are not occurring within the central
regions we probe here.  It is worth noting that in Paper I we identified a possible infalling
substructure to the north-east of NGC\,5044 at $\sim$1.4\,Mpc projected separation, but our stellar
population measurements do not extend that far and so we are unable to comment on possible trends at
these large radii.

While we do not see any obvious trends of stellar populations with radius, as Smith \etal do, we do
perhaps find a more subtle interpretation of galaxy evolution in the group environment through
analysis of emission-line properties.  Comparison of emission- and absorption-line metallicities has
allowed us to extract information regarding the previous star-formation episodes of galaxies through
the relation of nebular and stellar metallicities.  What we find is that the relative offset between
these two metallicity estimates is too large to be explained by either pure passive evolution of
galaxies, which can account for at most $\sim1$\,dex of offset, or ongoing star formation, which is
inconsistent with other observed galaxy properties.  The mass-dependence of our emission--absorption
metallicity offset instead is consistent with a scenario in which galaxies are being stripped of
their gas primarily through galaxy--IGM interactions, and therefore subsequent ISM enrichment is
having a larger effect on the observed emission-line metallicity.  We require a larger sample of
galaxies to confirm such an analysis, however this method provides a novel technique for probing
environmental effects on galaxy evolution. 

We turn, finally, to the question of NGC\,5044's evolution as a group and what information the
galaxy population holds in this regard.  The relative lack of radial trends in the stellar
population data limit the extent to which the gradual accretion scheme present in semi-analytic
models of group formation (e.g. de Lucia \etal 2006; Berrier \etal 2008) can be acting in the
NGC\,5044 group, at least without considerable mixing of the galaxy distribution with radius (e.g.
Ludlow \etal 2008).  The large peculiar velocity of NGC\,5044 itself relative to the group mean
($\sim200$\kms), possible presence of substructure on the outer regions of the group (see Paper I)
and recent detection of a ``sloshing'' type cold front in the group's X-ray distribution
(Gastaldello \etal 2008) suggest that the group is still growing and in the process of relaxation.
However, the group's Gaussian line-of-sight velocity distribution and relatively uniform X-ray
profiles argue against any recent, dynamically violent evolution of the group.

The work presented here represents the first targeted effort at constraining the stellar population
trends of group galaxies.  Further work is clearly needed, particularly on larger samples of
homogeneously selected galaxies, to better constrain the processes driving galaxy evolution in
groups.

\section*{Acknowledgements}

The authors would like to thank the anonymous referee for his/her helpful comments and suggestions.
We would also like to thank Marc Sarzi for his help in using {\sc gandalf} and Sergio Cellone for
providing us with his imaging and photometric measurements of dwarf galaxies in the NGC\,5044 group.

We acknowledge the analysis facilities provided by {\tiny IRAF}, which is distributed by the
National Optical Astronomy Observatories, which is operated by the Association of Universities for
Research in Astronomy, Inc., under cooperative agreement with the National Science Foundation.  This
publication makes use of data products from the Two Micron All Sky Survey, which is a joint project
of the University of Massachusetts and the Infrared Processing and Analysis Center/California
Institute of Technology, funded by the National Aeronautics and Space Administration and the
National Science Foundation.  We also thank the Australian Research Council for funding that
supported this work.

\begin{appendix}
\section{Index Errors}
\label{index_errors}

Here we show the error curves for all Lick/IDS indices as derived from AAOmega
observations.

\begin{figure}
\centering
\includegraphics[scale=0.95,angle=-90]{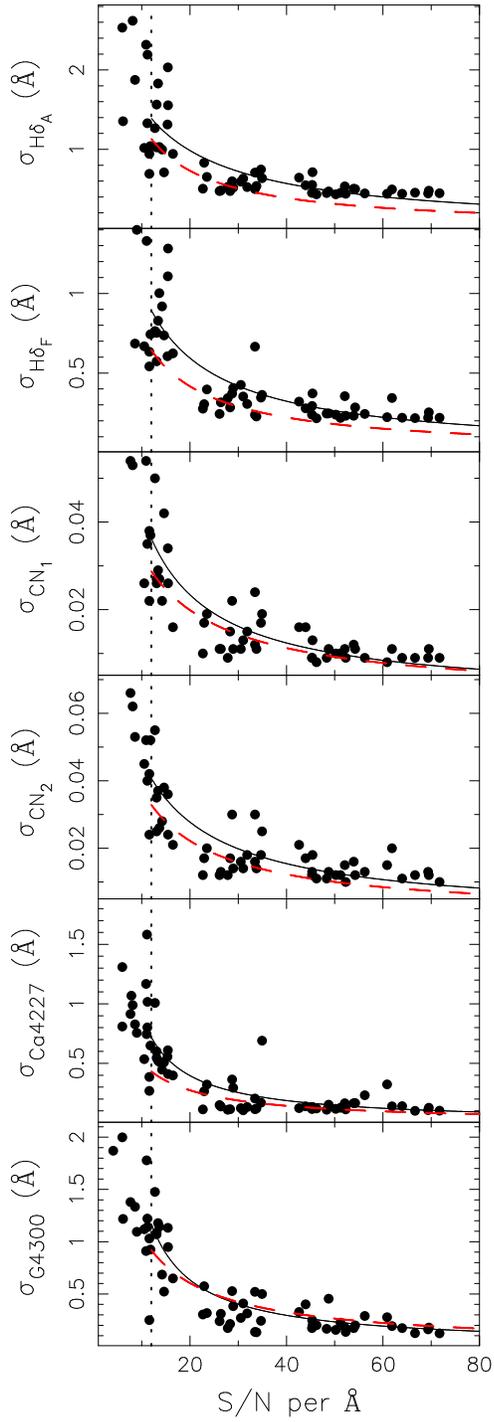}
\caption{Index Error as a function of signal to noise.  {\it Solid} (black) and {\it dashed} (red)
lines represent the best fit hyperbolic functions for these data and the 6dFGS DR1 data of P+08
respectively.  The vertical dotted line represents our adopted S/N cut of 12.} 
\label{ierr_1-6}
\end{figure}

\begin{figure}
\centering
\includegraphics[scale=0.95,angle=-90]{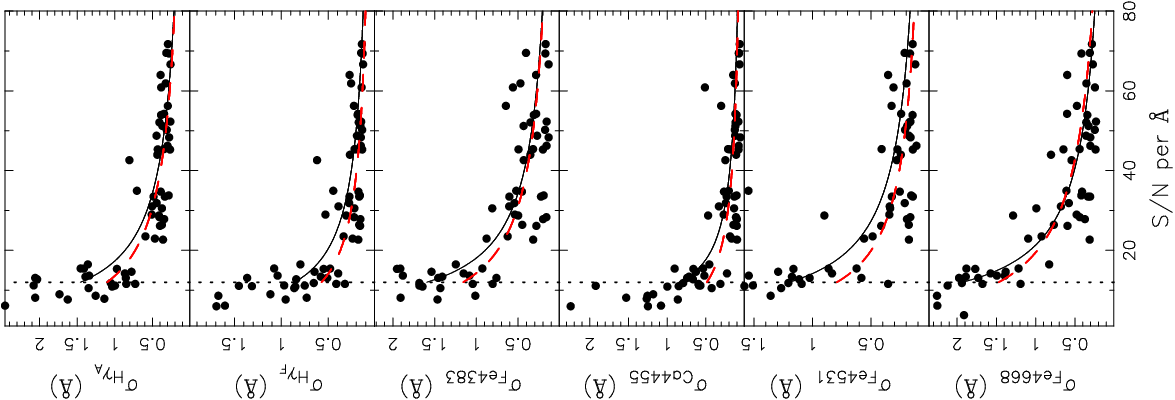}
\caption{Same as Fig. \ref{ierr_1-6}} 
\label{ierr_7-12}
\end{figure}

\begin{figure}
\centering
\includegraphics[scale=0.95,angle=-90]{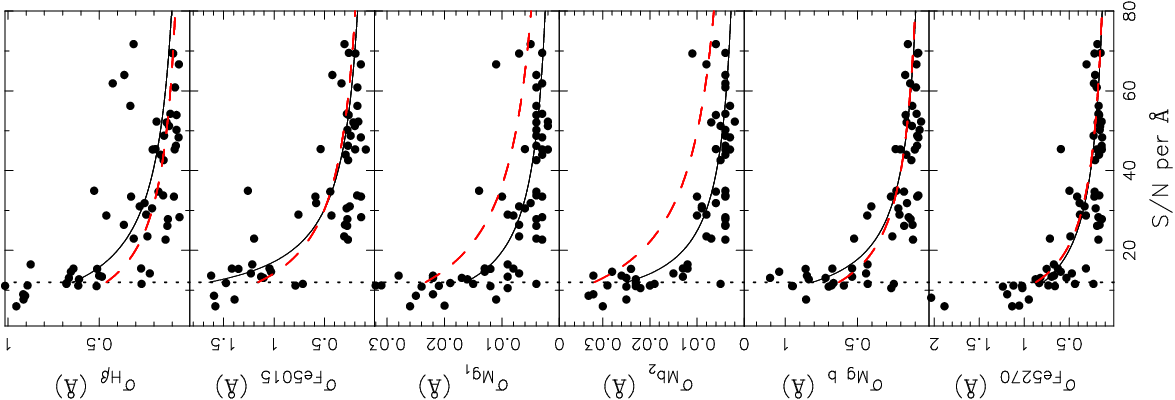}
\caption{Same as Fig. \ref{ierr_1-6}} 
\label{ierr_13-18}
\end{figure}

\begin{figure}
\centering
\includegraphics[scale=0.95,angle=-90]{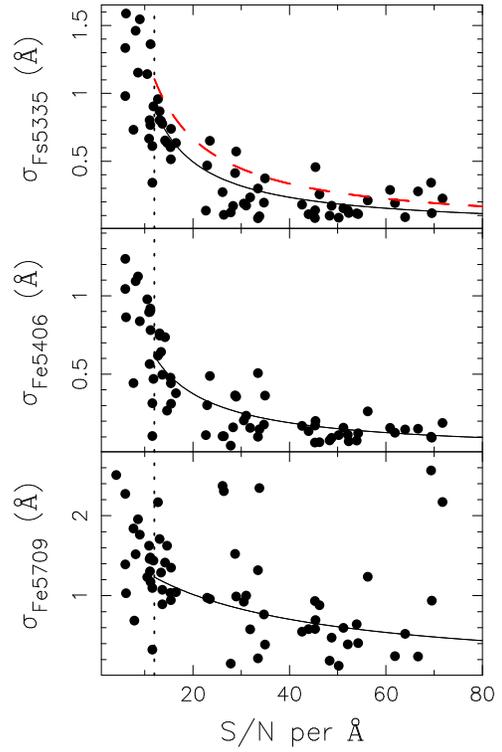}
\caption{Same as Fig. \ref{ierr_1-6}.  Fe5406 and Fe5709 indices were not measured in the 6dFGS DR1,
so their error curves are omitted.} 
\label{ierr_19-21}
\end{figure}

\clearpage

\section{Stellar population measurements}
\label{sp_app}

Here we include the derived stellar mass, M/L, age, metallicity and $\alpha$-element abundance
for all 67 galaxies in our spectral sample.

\begin{onecolumn}
\begin{center}
\begin{deluxetable}{lllccccc}
\tabletypesize{\footnotesize}
\tablecaption{Stellar population parameters of NGC\,5044 group galaxies.  Only those galaxies with S/N $>$ 12 are included (see text.)}
\tablewidth{0pt}
\tablehead{
\colhead{Galaxy Identifier} & \colhead{RA} & \colhead{DEC} & \colhead{$\log\,M_{\mathrm{*}}$} & \colhead{M/L$_{\mathrm{B}}$} & \colhead{$\log\,$age} & \colhead{[Fe/H]} & \colhead{[$\alpha$/Fe]}\\
\colhead{} & \colhead{(J2000)} & \colhead{(J2000)} &\colhead{} & \colhead{} & \colhead{} & \colhead{(dex)} & \colhead{(dex)} \\
\colhead{(1)} & \colhead{(2)} & \colhead{(3)} & \colhead{(4)} & \colhead{(5)} & \colhead{(6)} &\colhead{(7)} &\colhead{(8)} 
}
\startdata
FS90 003 & 13 12 26.35 & -15 47 52.30 & 9.3$^{+0.1}_{-0.1}$ & 0.64$^{+0.13}_{-0.12}$ & 0.20$\pm$0.08 & 0.03$\pm$0.06 & 0.30$\pm$0.08 \\
FS90 005 & 13 12 54.50 & -16 45 57.00 & 10.5$^{+0.1}_{-0.1}$ & 3.14$^{+0.15}_{-0.14}$ & 0.75$\pm$0.05 & 0.33$\pm$0.05 & 0.18$\pm$0.05 \\
FS90 009 & 13 13 5.49 & -16 28 41.72 & 9.9$^{+0.1}_{-0.1}$ & 2.98$^{+0.12}_{-0.18}$ & 0.70$\pm$0.05 & 0.25$\pm$0.05 & 0.06$\pm$0.05 \\
FS90 011 & 13 13 6.80 & -16 41 0.00 & 7.8$^{+0.1}_{-0.1}$ & 1.33$^{+0.13}_{-0.14}$ & 0.70$\pm$0.10 & -1.10$\pm$0.08 & -0.09$\pm$0.20 \\
FS90 015 & 13 13 12.48 & -16 7 50.15 & 9.9$^{+0.1}_{-0.1}$ & 2.99$^{+0.22}_{-0.20}$ & 0.82$\pm$0.05 & -0.05$\pm$0.05 & 0.03$\pm$0.05 \\
FS90 017 & 13 13 28.42 & -16 18 52.90 & 9.0$^{+0.1}_{-0.1}$ & 1.12$^{+0.17}_{-0.19}$ & 0.55$\pm$0.10 & -0.68$\pm$0.10 & 0.03$\pm$0.10 \\
FS90 018 & 13 13 32.24 & -17 4 43.40 & 9.7$^{+0.1}_{-0.1}$ & 1.49$^{+0.18}_{-0.13}$ & 0.45$\pm$0.07 & 0.07$\pm$0.07 & 0.03$\pm$0.06 \\
FS90 024 & 13 13 50.96 & -16 5 13.70 & 8.2$^{+0.1}_{-0.1}$ & 0.78$^{+0.15}_{-0.10}$ & 0.42$\pm$0.10 & -1.02$\pm$0.10 & -0.15$\pm$0.19 \\
FS90 027 & 13 13 54.15 & -16 29 27.40 & 9.9$^{+0.1}_{-0.1}$ & 2.46$^{+0.11}_{-0.20}$ & 0.70$\pm$0.05 & 0.05$\pm$0.05 & 0.09$\pm$0.05 \\
FS90 029 & 13 13 56.21 & -16 16 24.30 & 9.3$^{+0.0}_{-0.0}$ & 2.24$^{+0.10}_{-0.06}$ & 0.72$\pm$0.05 & -0.20$\pm$0.05 & 0.00$\pm$0.07 \\
FS90 030 & 13 13 59.52 & -16 23 3.80 & 9.3$^{+0.0}_{-0.0}$ & 3.00$^{+0.22}_{-0.27}$ & 1.07$\pm$0.05 & -0.62$\pm$0.05 & 0.15$\pm$0.08 \\
FS90 031 & 13 14 0.47 & -16 0 44.20 & 8.4$^{+0.1}_{-0.1}$ & 1.01$^{+0.11}_{-0.03}$ & 0.70$\pm$0.08 & -1.25$\pm$0.06 & 0.40$\pm$0.26 \\
FS90 032 & 13 14 3.22 & -16 7 23.20 & 10.3$^{+0.1}_{-0.1}$ & 5.10$^{+0.38}_{-0.35}$ & 1.07$\pm$0.05 & 0.15$\pm$0.05 & 0.24$\pm$0.05 \\
FS90 034 & 13 14 7.40 & -16 25 35.80 & 9.0$^{+0.0}_{-0.0}$ & 3.00$^{+0.14}_{-0.14}$ & 1.02$\pm$0.05 & -0.57$\pm$0.05 & 0.00$\pm$0.07 \\
FS90 038 & 13 14 13.24 & -16 19 29.00 & 8.7$^{+0.0}_{-0.0}$ & 3.43$^{+0.22}_{-0.21}$ & 1.02$\pm$0.05 & -0.30$\pm$0.05 & 0.12$\pm$0.06 \\
FS90 040 & 13 14 0.30 & -16 47 10.00 & 8.9$^{+0.2}_{-0.1}$ & 1.38$^{+0.29}_{-0.20}$ & 0.68$\pm$0.16 & -0.95$\pm$0.12 & -0.18$\pm$0.30 \\
FS90 042 & 13 14 17.36 & -16 26 19.50 & 9.3$^{+0.0}_{-0.0}$ & 2.81$^{+0.15}_{-0.14}$ & 0.97$\pm$0.05 & -0.45$\pm$0.05 & 0.09$\pm$0.06 \\
FS90 049 & 13 14 31.23 & -16 22 47.60 & 8.6$^{+0.1}_{-0.1}$ & 0.69$^{+0.11}_{-0.15}$ & 0.38$\pm$0.13 & -1.20$\pm$0.15 & -0.18$\pm$0.17 \\
FS90 050 & 13 14 34.86 & -16 29 28.90 & 8.8$^{+0.0}_{-0.1}$ & 0.84$^{+0.06}_{-0.09}$ & 0.38$\pm$0.08 & -0.42$\pm$0.08 & 0.12$\pm$0.06 \\
FS90 051 & 13 14 35.84 & -16 34 50.40 & 8.8$^{+0.0}_{-0.1}$ & 1.65$^{+0.08}_{-0.22}$ & 0.57$\pm$0.06 & -0.28$\pm$0.08 & 0.03$\pm$0.07 \\
FS90 053 & 13 14 39.22 & -15 50 36.00 & 8.3$^{+0.1}_{-0.0}$ & 0.74$^{+0.09}_{-0.06}$ & 0.45$\pm$0.07 & -1.18$\pm$0.08 & 0.06$\pm$0.10 \\
FS90 054 & 13 14 39.40 & -16 10 4.00 & 9.0$^{+0.1}_{-0.1}$ & 2.89$^{+0.35}_{-0.38}$ & 1.07$\pm$0.10 & -0.70$\pm$0.08 & 0.18$\pm$0.20 \\
FS90 058 & 13 14 43.61 & -15 51 34.40 & 7.7$^{+0.1}_{-0.1}$ & 1.00$^{+0.28}_{-0.26}$ & 0.60$\pm$0.20 & -1.07$\pm$0.16 & 0.15$\pm$0.25 \\
FS90 063 & 13 14 49.82 & -16 58 24.20 & 9.7$^{+0.0}_{-0.0}$ & 2.58$^{+0.19}_{-0.14}$ & 0.82$\pm$0.05 & -0.17$\pm$0.05 & 0.09$\pm$0.06 \\
FS90 064 & 13 14 49.23 & -16 29 33.69 & 9.7$^{+0.1}_{-0.1}$ & 1.06$^{+0.10}_{-0.11}$ & 0.30$\pm$0.05 & 0.17$\pm$0.06 & 0.12$\pm$0.05 \\
FS90 068 & 13 14 59.37 & -16 35 25.10 & 10.6$^{+0.1}_{-0.1}$ & 3.44$^{+0.24}_{-0.21}$ & 0.88$\pm$0.05 & 0.17$\pm$0.05 & 0.21$\pm$0.05 \\
FS90 072 & 13 15 2.13 & -15 57 6.50 & 9.3$^{+0.1}_{-0.1}$ & 0.40$^{+0.02}_{-0.04}$ & 0.00$\pm$0.06 & -0.12$\pm$0.05 & 0.06$\pm$0.06 \\
FS90 075 & 13 15 4.08 & -16 23 40.20 & 9.1$^{+0.0}_{-0.0}$ & 2.18$^{+0.11}_{-0.04}$ & 0.75$\pm$0.05 & -0.17$\pm$0.05 & 0.12$\pm$0.06 \\
FS90 078 & 13 15 10.54 & -16 14 19.55 & 9.8$^{+0.1}_{-0.1}$ & 3.52$^{+0.24}_{-0.22}$ & 1.02$\pm$0.05 & -0.12$\pm$0.05 & 0.27$\pm$0.06 \\
FS90 079 & 13 15 12.71 & -16 29 57.10 & 9.2$^{+0.0}_{-0.0}$ & 3.21$^{+0.22}_{-0.15}$ & 0.97$\pm$0.05 & -0.28$\pm$0.05 & 0.09$\pm$0.06 \\
FS90 081 & 13 15 14.30 & -15 50 59.00 & 8.0$^{+0.2}_{-0.2}$ & 2.55$^{+0.56}_{-0.54}$ & 1.07$\pm$0.16 & -0.97$\pm$0.09 & 0.21$\pm$0.23 \\
FS90 082 & 13 15 17.57 & -16 29 10.19 & 9.4$^{+0.1}_{-0.0}$ & 1.37$^{+0.15}_{-0.12}$ & 0.45$\pm$0.07 & -0.05$\pm$0.07 & 0.03$\pm$0.06 \\
FS90 084 & 13 15 23.97 & -16 23 7.90 & 11.3$^{+0.1}_{-0.1}$ & 6.85$^{+0.49}_{-0.41}$ & 1.18$\pm$0.05 & 0.25$\pm$0.05 & 0.24$\pm$0.05 \\
FS90 094 & 13 15 32.04 & -16 28 51.10 & 9.5$^{+0.0}_{-0.0}$ & 3.63$^{+0.20}_{-0.20}$ & 1.00$\pm$0.05 & -0.23$\pm$0.05 & 0.06$\pm$0.06 \\
FS90 095 & 13 15 32.40 & -16 38 12.00 & 8.3$^{+0.1}_{-0.1}$ & 2.40$^{+0.11}_{-0.14}$ & 1.07$\pm$0.08 & -1.02$\pm$0.05 & 0.30$\pm$0.20 \\
FS90 096 & 13 15 32.70 & -16 50 35.00 & 7.9$^{+0.2}_{-0.2}$ & 1.16$^{+0.37}_{-0.27}$ & 0.80$\pm$0.16 & -1.50$\pm$0.12 & 0.56$\pm$0.32 \\
FS90 100 & 13 15 45.12 & -16 19 36.63 & 10.2$^{+0.1}_{-0.1}$ & 5.01$^{+0.38}_{-0.35}$ & 1.07$\pm$0.05 & 0.10$\pm$0.05 & 0.21$\pm$0.05 \\
FS90 102 & 13 15 48.52 & -16 31 8.00 & 10.0$^{+0.1}_{-0.1}$ & 1.59$^{+0.08}_{-0.19}$ & 0.60$\pm$0.06 & -0.07$\pm$0.07 & 0.34$\pm$0.07 \\
FS90 107 & 13 15 59.30 & -16 23 49.80 & 10.3$^{+0.1}_{-0.1}$ & 3.73$^{+0.30}_{-0.26}$ & 0.93$\pm$0.05 & 0.20$\pm$0.05 & 0.27$\pm$0.05 \\
FS90 108 & 13 15 59.90 & -16 10 32.00 & 8.6$^{+0.1}_{-0.1}$ & 2.78$^{+0.32}_{-0.33}$ & 1.07$\pm$0.09 & -0.85$\pm$0.05 & 0.12$\pm$0.14 \\
FS90 117 & 13 16 23.11 & -16 8 11.35 & 9.4$^{+0.0}_{-0.0}$ & 2.83$^{+0.17}_{-0.15}$ & 0.95$\pm$0.05 & -0.30$\pm$0.05 & 0.18$\pm$0.06 \\
FS90 123 & 13 16 35.62 & -16 26 5.70 & 8.5$^{+0.1}_{-0.1}$ & 3.45$^{+0.24}_{-0.22}$ & 1.15$\pm$0.06 & -0.68$\pm$0.05 & 0.09$\pm$0.10 \\
FS90 127 & 13 16 39.40 & -16 57 3.00 & 8.4$^{+0.1}_{-0.1}$ & 3.41$^{+0.25}_{-0.22}$ & 1.02$\pm$0.07 & -0.70$\pm$0.06 & -0.30$\pm$0.10 \\
FS90 133 & 13 16 55.43 & -16 26 32.20 & 7.7$^{+0.1}_{-0.1}$ & 0.82$^{+0.09}_{-0.10}$ & 0.53$\pm$0.09 & -0.93$\pm$0.08 & 0.46$\pm$0.17 \\
FS90 134 & 13 16 56.23 & -16 35 34.70 & 9.1$^{+0.1}_{-0.1}$ & 0.75$^{+0.08}_{-0.06}$ & 0.47$\pm$0.07 & -1.25$\pm$0.08 & 0.06$\pm$0.13 \\
FS90 135 & 13 16 54.40 & -16 3 18.00 & 8.2$^{+0.1}_{-0.1}$ & 2.47$^{+0.18}_{-0.21}$ & 1.07$\pm$0.09 & -0.85$\pm$0.06 & 0.42$\pm$0.23 \\
FS90 137 & 13 16 58.49 & -16 38 5.50 & 10.5$^{+0.1}_{-0.1}$ & 0.68$^{+0.10}_{-0.14}$ & 0.12$\pm$0.07 & 0.10$\pm$0.06 & 0.00$\pm$0.06 \\
FS90 138 & 13 17 3.10 & -16 35 39.00 & 8.0$^{+0.1}_{-0.1}$ & 2.04$^{+0.17}_{-0.21}$ & 1.02$\pm$0.10 & -1.02$\pm$0.05 & 0.46$\pm$0.23 \\
FS90 144 & 13 17 10.79 & -16 13 46.74 & 9.9$^{+0.1}_{-0.1}$ & 3.49$^{+0.23}_{-0.22}$ & 1.05$\pm$0.05 & -0.17$\pm$0.05 & 0.30$\pm$0.05 \\
FS90 153 & 13 17 36.38 & -16 32 25.40 & 9.5$^{+0.0}_{-0.0}$ & 2.60$^{+0.14}_{-0.15}$ & 0.88$\pm$0.05 & -0.33$\pm$0.05 & 0.06$\pm$0.08 \\
FS90 158 & 13 18 9.04 & -16 58 15.60 & 8.9$^{+0.1}_{-0.1}$ & 2.92$^{+0.19}_{-0.25}$ & 1.10$\pm$0.06 & -0.97$\pm$0.05 & 0.00$\pm$0.11 \\
FS90 161 & 13 18 17.80 & -16 38 25.00 & 8.0$^{+0.2}_{-0.1}$ & 1.91$^{+0.77}_{-0.40}$ & 1.02$\pm$0.20 & -1.20$\pm$0.07 & 0.42$\pm$0.37 \\
2MASXJ13102493-1655578 & 13 10 24.96 & -16 55 57.50 & 9.8$^{+0.2}_{-0.2}$ & 1.41$^{+0.32}_{-0.29}$ & 0.70$\pm$0.15 & -0.57$\pm$0.13 & 0.32$\pm$0.15 \\
2MASXJ13114576-1915421 & 13 11 45.77 & -19 15 42.30 & \ldots & 4.11$^{+0.29}_{-0.30}$ & 1.00$\pm$0.06 & -0.05$\pm$0.05 & 0.09$\pm$0.07 \\
2MASXJ13115849-1644541 & 13 11 58.49 & -16 44 54.10 & 9.6$^{+0.1}_{-0.1}$ & 2.98$^{+0.12}_{-0.12}$ & 1.00$\pm$0.06 & -0.53$\pm$0.05 & 0.00$\pm$0.12 \\
2MASXJ13125449-1645571 (FS90 005)& 13 12 54.50 & -16 45 57.00 & 10.6$^{+0.1}_{-0.1}$ & 4.07$^{+0.25}_{-0.26}$ & 1.02$\pm$0.06 & -0.10$\pm$0.05 & 0.12$\pm$0.07 \\
2MASXJ13130549-1628411 (FS90 009)& 13 13 5.49 & -16 28 41.72 & 9.9$^{+0.1}_{-0.1}$ & 3.23$^{+0.28}_{-0.30}$ & 0.85$\pm$0.08 & -0.07$\pm$0.06 & -0.03$\pm$0.10 \\
2MASXJ13143041-1732009 & 13 14 30.42 & -17 32 0.90 & 9.4$^{+0.1}_{-0.1}$ & 1.83$^{+0.18}_{-0.12}$ & 0.90$\pm$0.07 & -1.40$\pm$0.05 & -0.30$\pm$0.14 \\
2MASXJ13143485-1629289 (FS90 050)& 13 14 34.86 & -16 29 28.90 & 8.5$^{+0.5}_{-0.2}$ & 0.41$^{+0.47}_{-0.13}$ & 0.10$\pm$0.29 & -0.57$\pm$0.20 & -0.30$\pm$0.12 \\
2MASXJ13150409-1623391 (FS90 075)& 13 15 4.08 & -16 23 40.20 & 9.4$^{+0.1}_{-0.1}$ & 3.83$^{+0.29}_{-0.27}$ & 0.97$\pm$0.07 & -0.35$\pm$0.05 & -0.21$\pm$0.09 \\
2MASXJ13153203-1628509 (FS90 094)& 13 15 32.04 & -16 28 51.10 & 9.1$^{+0.2}_{-0.2}$ & 1.64$^{+0.38}_{-0.34}$ & 0.45$\pm$0.14 & 0.00$\pm$0.14 & -0.18$\pm$0.12 \\
2MASXJ13164875-1620397 & 13 16 48.75 & -16 20 39.70 & 9.5$^{+0.1}_{-0.1}$ & 2.93$^{+0.22}_{-0.25}$ & 1.07$\pm$0.07 & -0.75$\pm$0.05 & 0.09$\pm$0.16 \\
2MASXJ13165533-1756417 & 13 16 55.35 & -17 56 42.00 & 9.5$^{+0.1}_{-0.1}$ & 2.41$^{+0.28}_{-0.28}$ & 0.97$\pm$0.08 & -0.88$\pm$0.05 & -0.06$\pm$0.14 \\
2MASXJ13165624-1635347 (FS90 134)& 13 16 56.23 & -16 35 34.70 & 9.6$^{+0.1}_{-0.1}$ & 2.50$^{+0.22}_{-0.24}$ & 1.07$\pm$0.09 & -1.50$\pm$0.06 & -0.30$\pm$0.18 \\
2MASXJ13171239-1715162 & 13 17 12.40 & -17 15 16.10 & 9.7$^{+0.1}_{-0.1}$ & 1.34$^{+0.18}_{-0.13}$ & 0.62$\pm$0.10 & -0.78$\pm$0.08 & -0.15$\pm$0.11 \\
2MASXJ13184125-1904476 & 13 18 41.26 & -19 4 47.70 & 8.9$^{+0.2}_{-0.2}$ & 0.39$^{+0.18}_{-0.12}$ & -0.05$\pm$0.18 & -0.10$\pm$0.07 & -0.06$\pm$0.15 \\
2MASXJ13185909-1835167 & 13 18 59.09 & -18 35 16.70 & 10.1$^{+0.1}_{-0.1}$ & 3.21$^{+0.24}_{-0.25}$ & 0.85$\pm$0.06 & 0.03$\pm$0.05 & 0.09$\pm$0.07 \\
\enddata
\label{sp_tab}
\end{deluxetable}
\end{center}
\end{onecolumn}
\clearpage

\end{appendix}


\begin{thebibliography}{999}
\bibitem[am99]{am99}Abadi M. G., Moore B., Bower R. G., 1999, MNRAS, 308, 947
\bibitem[ap01]{ap01}Allende Prieto C. A., Lambert D. L., Asplund M., 2001, ApJ, 556, 63
\bibitem[ba06]{ba06}Baldry I. K., Balogh M. L., Bower R. G., Glazebrook K., Nichol R. C., Bamford S. P., Budavari T., 2006, MNRAS, 373, 469
\bibitem[bg08]{bg08}Baldry I. K., Glazebrook K., Driver S. P., 2008, MNRAS, 388, 945
\bibitem[bp81]{bp81}Baldwin J. A., Phillips M. M., Terlevich R., 1981, PASP, 93, 5
\bibitem[ba07]{ba07}Balestra I., Tozzi P., Ettori S., Rosati P., Borgani S., Mainieri V., Norman C., Viola M., 2007, A\&A, 462, 429
\bibitem[bm00]{bm00}Balogh M. L., Morris S. L., 2000, MNRAS, 318, 703 
\bibitem[bh92]{bh92}Barnes J. E., Hernquist L., 1992, ARA\&A, 30, 705 
\bibitem[be04]{be04}Beasley M. A., Brodie J. P., Strader J., Forbes D. A., Proctor R. N., Barmby P., Huchra J. P., 2004, AJ, 128, 1623 
\bibitem[bm04]{bm04}B\"ohringer H., Matsushita K., Churazov E., Finoguenov A., Ikebe Y., 2004, A\&A, 416, 21
\bibitem[bd07]{bd07}Bell E. F., de Jong R. S., 2001, ApJ, 550, 212
\bibitem[bl04]{bl04}Bell E. F., et al., 2004, ApJ, 608, 752
\bibitem[be08]{be08}Berrier J. C., Stewart K. R., Bullock J. S., Purcell C. W., Barton E. J., Wechsler R. H., 2008, preprint (astro-ph/0804.0426)
\bibitem[bl03]{bl03}Blanton M. R., et al., 2003, ApJ, 592, 819 
\bibitem[bl05]{bl05}Blanton M. R., Eisenstein D., Hogg D. W., Schlegel D. J., Brinkmann J., 2005, ApJ, 629, 143
\bibitem[bb07]{bb07}Blanton, M. R., Berlind A. A., 2007, ApJ, 664, 791
\bibitem[br07]{br07}Brooks A. M., Governato F., Booth C. M., Willman B., Gardner J. P., Wadsley J., Stinson G., Quinn T., 2007, ApJ, 655, 17
\bibitem[bn07]{bn07}Brown M. J. I., Dey A., Jannuzi B. T., Brand K., Benson A. J., Brodwin M., Croton D. J., Eisenhardt P. T., 2007, ApJ, 654, 858
\bibitem[bp07]{bp07}Brough S., Proctor R., Forbes D. A., Couch W. J., Collins C. A., Burke D. J., Mann R. G., 2007, MNRAS, 378, 1507
\bibitem[bc03]{bc03}Bruzual G., Charlot S., 2003, MNRAS, 344, 1000 (BC03)
\bibitem[bu84]{bu84}Burstein D., Faber S. M., Gaskell C. M., Krumm N., 1984, ApJ, 287, 586
\bibitem[bu06]{bu06}Bundy K., et al., 2006, ApJ, 651, 120
\bibitem[bu04]{bu04}Buote D. A., Brighenti F., Mathews W. G., 2004, ApJ, 607, 91
\bibitem[ca93]{ca93}Caldwell N., Rose J. A., Sharples R. M., Ellis R. S., Bower R. G., 1993, AJ, 106, 473
\bibitem[cr03]{cr03}Caldwell N., Rose J. A., Concannon K. D., 2003, AJ, 125, 2891
\bibitem[cc89]{cc89}Cardelli J. A., Clayton G. C., Mathis J. S., 1989, ApJ, 345, 245
\bibitem[ca02]{ca02}Carter D., \etal, 2002, ApJ, 567, 772
\bibitem[ce04]{ce04}Cappellari M., Emsellem E., 2004, PASP, 116, 138
\bibitem[ce99]{ce99}Cellone S. A., 1999, A\&A, 345, 403
\bibitem[cb01]{cb01}Cellone S. A., Buzzoni A., 2001, A\&A, 369, 742
\bibitem[cb05]{cb05}Cellone S. A., Buzzoni A., 2005, MNRAS, 356, 41
\bibitem[ch03]{ch03}Chabrier G., 2003, PASP, 115, 763
\bibitem[co08]{co08}Conroy C., Gunn J. E., White M., 2008, preprint (astro-ph/0809.4261)
\bibitem[ct08]{ct08}Cooper M. C., Tremonti C. A., Newman J. A., Zabludoff A. I., 2008, arXiv:0805.0308
\bibitem[dl06]{dl06}De Lucia G., Springel V., White S. D. M., Croton D., Kauffmann G., 2006, MNRAS, 366, 499
\bibitem[dm06]{dm06}Di Matteo P., Combes F., Melchior A.-L., Semelin B., 2007, A\&A, 468, 61
\bibitem[de80]{de80}Dressler A., 1980, ApJ, 236, 351
\bibitem[el08]{el08}Ellison S. L., Patton D. R., Simard L., McConnachie A. W., 2008, AJ, 135, 1877
\bibitem[fa05]{fa05}Faber S. M., et al., 2007, ApJ, 665, 265 
\bibitem[fm05]{fm05}Faltenbacher A., Mathews W. G., 2005, MNRAS, 362, 498 (FM05)
\bibitem[fs81]{fs81}Farouki R., Shapiro S. L., 1981, ApJ, 243, 32
\bibitem[fs90]{fs90}Ferguson H. C., Sandage A., 1990, AJ, 100, 1 (FS90)
\bibitem[fi00]{fi00}Finoguenov A., David L. P., Ponman T. J., 2000, ApJ, 544, 188
\bibitem[ga06]{ga06}Gallazzi A., Charlot S., Brinchmann J., White S. D. M., 2006, MNRAS, 370, 1106
\bibitem[ga04]{ga04}Gao L., White S. D. M., Jenkins A., Stoehr F., Springel V., 2004, MNRAS, 255, 819
\bibitem[gi97]{gi97}Gibson B. K., Matteucci F., 1997, ApJ, 475, 435
\bibitem[go03]{go03}G\'omez P. L., et al., 2003, ApJ, 584, 210
\bibitem[go07]{go07}Gonzalez A. H., Zaritsky D., Zabludoff A. I., 2007, ApJ, 666, 147
\bibitem[gt03]{gt03}Goto T., Yamauchi C., Fujita Y., Okamura S., Sekiguchi M., Smail I., Bernardi M., G\'omez P. L., 2003, MNRAS, 346, 601
\bibitem[ga08]{ga08}Gastaldello F., Buote D. A., Temi P., Brighenti F., Mathews W. G., Ettori S., 2008, arXiv:0807.3526
\bibitem[gg72]{gg72}Gunn J. E., Gott J. R., 1972, ApJ, 176, 1
\bibitem[gu97]{gu97}Guzman R., Gallego J., Koo D. C., Phillips A. C., Lowenthal J. D., Faber S. M., Illingworth G. D., Vogt N. P., 1997, ApJ, 489, 559
\bibitem[he06]{he06}Hester J. A., 2006, ApJ, 647, 910
\bibitem[ho02]{ho02}Hogg D. W., et al., 2002, AJ, 124, 646
\bibitem[ja00]{ja00}Jarrett T. H., Chester T., Cutri R., Schneider S., Skrutskie M., Huchra J. P., 2000, AJ, 119, 2498 
\bibitem[je03]{je03}Jensen J. B., Tonry J. L., Barris B. J., Thompson R. I., Liu M. C., Rieke M. J., Ajhar E. A., Blakeslee J. P., 2003, ApJ, 583, 712
\bibitem[ji07]{ji07}Jimenez R., Bernardi M., Haiman Z., Panter B., Heavens A. F., 2007, ApJ, 669, 947
\bibitem[jo04]{jo04}Jones D. H., et al., 2004, MNRAS, 355, 747
\bibitem[jo05]{jo05}Jones D. H., Saunders W., Read M., Colless M., 2005, PASA, 22, 277
\bibitem[jf95]{jf95}J\o rgensen I., Franx M., Kj\ae rgaard P., 1995, MNRAS, 276, 1341
\bibitem[ka96]{ka96}Kauffmann G., 1996, MNRAS, 281, 487
\bibitem[ka03]{ka03}Kauffmann G., et al., 2003, MNRAS, 346, 1055
\bibitem[ka04]{ka04}Kauffmann G., White S. D. M., Heckman T. M., M\'enard B., Brinchmann J., Charlot S., Tremonti C., Brinkmann J., 2004, MNRAS, 353, 713
\bibitem[ka08]{ka08}Kawata D., Mulchaey J. S., 2008, ApJ, 672, 103
\bibitem[ke01]{ke01}Kewley L. J., Dopita M. A., Sutherland R. S., Heisler C. A., Trevena J., 2001, ApJ, 556, 121
\bibitem[ke04]{ke04}Kewley L. J., Geller M. J., JAnsen R. A., 2004, AJ, 127, 2002 
\bibitem[kd02]{kd02}Kewley L. J., Dopita M. A., 2002, ApJS, 142, 35
\bibitem[ko03]{ko03}Kobulnicky H. A., et al., 2003, ApJ, 599, 1006 
\bibitem[kk04]{kk04}Kobulnicky H. A., Kewley L. J., 2004, ApJ, 617, 240
\bibitem[kp03]{kp03}Kobulnicky H. A., Phillips A. C., 2003, ApJ, 599, 1031
\bibitem[ko08]{ko08}Komatsu E., \etal, 2008, preprint (astro-ph/0803.0547)
\bibitem[km05]{km05}Korn A. J., Maraston C., Thomas D., 2005, A\&A, 438, 685
\bibitem[la07]{la07}Lamareille F., Contini T., Charlot S., Brinchmann J., 2007, in Combes F., Palous J., ed., IAU Symposium 235, Galaxy Evolution Across the Hubble Time, Cambridge University Press, p. 408
\bibitem[lt80]{lt80}Larson R. B., Tinsley B. M., Caldwell C. N., 1980, ApJ, 237, 692
\bibitem[le02]{le02}Lewis I., et al., 2002, MNRAS, 334, 673
\bibitem[li07]{li07}Lin L., et al., 2007, ApJ, 660, 51
\bibitem[lu08]{lu08}Ludlow A. D., Navarro J. F., Springel V., Jenkins A., Frenk C. S., Helmi A., 2008 preprint (astro-ph/0801.1127v1)
\bibitem[ma05]{ma05}Machacek M. E., Nulsen P., Stirbat L., Jones C., Forman W. R., 2005, ApJ, 630, 280
\bibitem[ma08]{ma08}Maiolino R., 2008, A\&A, 488, 463
\bibitem[m+04]{m+04}Mathews W. G., Chomiuk L., Brighenti F., Buote D. A., 2004, ApJ, 616, 745 (M+04)
\bibitem[mc08]{mc08}McIntosh D. H., Guo Y., Hertzberg J., Katz N., Mo H. J., van den Bosch F. C., Yang X., 2008, MNRAS, 388, 1537
\bibitem[me07]{me07}Mendel J. T., Proctor R. N., Forbes D. A., 2007, MNRAS, 379, 1618
\bibitem[me08]{me08}Mendel J. T., Proctor R. N., Forbes D. A., Brough S., 2008, MNRAS, 389, 749 (Paper I)
\bibitem[mo01]{mo01}Mobasher B., \etal, 2001, ApJS, 137, 279
\bibitem[mo96]{mo96}Moore B., Katz N., Lake G., Dressler A., Oemler A., 1996, Nature, 379, 613
\bibitem[mo08]{mo08}Mouchine M., gibson B. K., Renda A., Kawata D., 2008, preprint (astro-ph/0801.2476)
\bibitem[mo03]{mo03}Moretti A., Portinari L., Chiosi C., 2003, A\&A, 408, 431
\bibitem[na95]{na95}Navarro, J. F., Frenk C. S., White S. D. M., 1995, MNRAS, 275, 56
\bibitem[na96]{na96}--------, 1996, ApJ, 462, 563
\bibitem[na97]{na97}--------, 1997, ApJ, 490, 493
\bibitem[nw83]{nw83}Negroponte J., White S. D. M., 1983, MNRAS, 205, 1009
\bibitem[nu82]{nu82}Nulsen P. E. J., 1982, MNRAS, 198, 1007
\bibitem[oe74]{oe74}Oemler A., 1974, ApJ, 194, 1
\bibitem[om08]{om08}O'mill A. L., Padilla N., Lambas D. G., 2008, MNRAS, 389, 1763 
\bibitem[os89]{os89}Osterbrock D. E., 1989, Astrophysics of gaseous nebulae and active galactic nuclei, University Science Books
\bibitem[pa00]{pa00}Paturel G., Fang Y., Petit C., Garnier R., Rousseau J., 2000, A\&AS, 146, 19
\bibitem[pa05]{pa05}Paturel G., Vauglin I., Petit C., Borsenberger J., Epchtein N., Fouqu\'{e} P., Mamon G., 2005, A\&A, 430, 751
\bibitem[pe07]{pe07}Peletier R. F., 2007, MNRAS, 379, 445
\bibitem[pi06]{pi06}Pierce M. et al., 2006, MNRAS, 366, 1253
\bibitem[po04]{po04}Poggianti B. M., Bridges T. J., Komiyama Y., Yagi M., Carter D., Mobasher B., Okamura S., Kashikawa N., 2004, ApJ, 601, 197
\bibitem[po06]{po06}Poggianti B. M., et al., 2006, ApJ, 642, 188
\bibitem[pf04]{pf04}Proctor R. N., Forbes D. A., Beasley M. A., 2004, MNRAS, 355, 1327
\bibitem[p+08]{p+08}Proctor R. N., Lah P., Forbes D. A., Colless M., Couch W., 2008, MNRAS, 386, 1781 (P+08)
\bibitem[ps02]{ps02}Proctor R. N., Sansom A. A., 2002, MNRAS, 333, 517
\bibitem[qm00]{qm00}Quilis V., Moore B., Bower R., 2000, Science, 288, 1617
\bibitem[rp06]{rp06}Rasmussen J., Ponman T. J., Mulchaey J. S., 2006, MNRAS, 370, 453
\bibitem[rp07]{rp07}Rasmussen J., Ponman T. J., 2007, MNRAS, 380, 1554
\bibitem[rp04]{rp04}Rickes M. G., Pastoriza M. G., Bonatto Ch., 2004, A\&A, 419, 449
\bibitem[sb06]{sb06}S\'anchez-Bl\'azquez P., et al., 2006, MNRAS, 371, 703
\bibitem[sb07]{sb07}S\'anchez-Bl\'azquez P., Forbes D. A., Strader J., Brodie J., Proctor R., 2007, MNRAS, 377, 759
\bibitem[sa98]{sa98}Sansom A. E., Proctor R. N., 1998, MNRAS, 297, 953
\bibitem[sa06]{sa06}Sarzi M., et al., 2006, MNRAS, 366, 1151
\bibitem[sa05]{sa05}Savaglio S., et al., 2005, ApJ, 635, 260
\bibitem[si04]{si04}Sivakoff G. R., Sarazin C. L., Carlin J. L., 2004, ApJ, 617, 262
\bibitem[sm07]{sm07}Smith R. J., Lucey J. R., Hudson M. J., 2007, MNRAS, 381, 1035
\bibitem[sm08]{sm08}Smith R. J., et al., 2008a, MNRAS, 386, 96
\bibitem[sm0b]{sm0b}Smith R. J., Lucey J. R., Hudson M. H., Allanson S. P., Bridges T. J., Hornschemeier A. E., Marzke R. O., Miller N. A., 2008b, arXiv:0810.5558
\bibitem[sp99]{sp99}Somerville R. S., Primack J. R., 1999, MNRAS, 310, 1087
\bibitem[sp08]{sp08}Spolaor M., Forbes D. A., Proctor R. N., Hau G. K. T., Brough S., 2008, MNRAS, 385, 675
\bibitem[sp05]{sp05}Springel V., et al., 2005, Nature, 435, 629
\bibitem[tf02]{tf02}Terlevich A. I., Forbes D. A., 2002, MNRAS, 330 547
\bibitem[te07]{te07}Temi P., Brighenti F., Mathews W. G., 2007, ApJ, 666, 222
\bibitem[tm03]{tm03}Thomas D., Maraston C., Bender R., 2003, MNRAS, 339, 897
\bibitem[th05]{th05}Thomas D., Maraston C., Bender R., Mendes de Oliveira C., 2005, ApJ, 621, 673
\bibitem[to01]{to01}Tonry J. L., Dressler A., Blakeslee J. P., Ajhar E. A., Fletcher A. B., Luppino G. A., Metzger M. R., Moore C. B., 2001, ApJ, 546, 681
\bibitem[tt72]{tt72}Toomre A., Toomre J., 1972, ApJ, 178, 623
\bibitem[to03]{to03}Tozzi P., Rosati P., Ettori S., Borgani S., Mainieri V., Norman C., 2003, ApJ, 593, 705
\bibitem[tr98]{tr98}Trager S. C., Worthey G., Faber S. M., Burstein D., Gonz\'alez J.J., 1998, ApJS, 116, 1
\bibitem[tr00]{tr00}Trager S. C., Faber S. M., Worthey G., Gonz\'alex J. J., 2000, AJ, 120, 165
\bibitem[tr08]{tr08}Trager S. C., Faber S. M., Dressler A., 2008, MNRAS, 386, 715
\bibitem[tr04]{tr04}Tremonti C. A., et al., 2004, ApJ, 613, 898
\bibitem[tr94]{tr94}Trentham N., 1994, Nature, 372, 157
\bibitem[vd08]{vd08}van den Bosch F. C., Pasquali A., Yang X., Mo H. J., Weinmann S., McIntosh D. H., Aquino D., 2008, arXiv:0805.0002 
\bibitem[vd07]{vd07}van den Bosch F. C., \etal, 2007, MNRAS, 376, 841
\bibitem[vd93]{vd93}van der Marel R. P., Franx M., 1993, ApJ, 407, 525
\bibitem[wc08]{wv08}Woo J., Courteau S., Dekel A., 2008 
\bibitem[wo97]{wo97}Worthey G., Ottaviani D. L., 1997, ApJS, 111, 377
\end{thebibliography}
\end{document}